\documentclass[oneside,a4paper,11pt,shownumbers,manualsort]{article}
\usepackage{latexsym}
\usepackage{euscript}
\usepackage{epsfig,amsmath,amssymb}
\usepackage{slashed}
 
\renewcommand\({\left(}
\renewcommand\){\right)}
\renewcommand\[{\left[}
\renewcommand\]{\right]}

\newcommand{\dd}{{\rm d}}
\newcommand{\e}{{\rm e}}
\def\be{\begin{equation}}
\def\ee{\end{equation}}
\def\bea{\begin{eqnarray}}
\def\eea{\end{eqnarray}}

\newcommand\TeV{\,\mbox{TeV}}
\newcommand\GeV{\,\mbox{GeV}}

\newcommand\mpl{m_{\rm p}}

\newcommand\mcO{\mathcal O}
\newcommand\mcN{\mathcal N}

\long\def\symbolfootnote[#1]#2{\begingroup%
\def\thefootnote{\fnsymbol{footnote}}\footnote[#1]{#2}\endgroup} 

 \large\normalsize

\begin{document}
%\draft

\begin{center}

{\Large \bf Enlarging the parameter space of standard hybrid inflation}

\vspace*{7mm} {\ Rachel Jeannerot$^{a}$
\symbolfootnote[1]{{E-mail:jeannerot@lorentz.leidenuniv.nl}}
and Marieke Postma$^{b}$}
\symbolfootnote[2]{E-mail:mpostma@nikhef.nl}
\vspace*{.25cm}

${}^{a)}${\it Instituut-Lorentz for Theoretical Physics,
Niels Bohrweg 2, 2333 CA Leiden, The Netherlands}\\
\vspace*{.1cm} 
${}^{b)}${\it NIKHEF, Kruislaan 409, 1098 SJ Amsterdam,
The Netherlands}

%\vspace*{.3cm}
\end{center}

\begin{abstract}
We show that the parameter space for F-term inflation which predict
the formation of cosmic strings is larger than previously
estimated. Firstly, because realistic embeddings in GUT theories alter
the standard scenerio, making the inflationary potential less steep.
Secondly, the strings which form at the end of inflation are not
necessarily topologically stable down to low scales. In shifted and
smooth inflation strings do not form at all.  We also discuss D-term
inflation; here the possibilities are much more limited to enlargen
paramer space.
\end{abstract}

\section{Introduction}

Hybrid inflation is perhaps the best particle physics motivated model
of inflation~\cite{hybrid}. In the standard model for hybrid inflation
in supersymmetric theories, inflation ends in a phase transition
during which spontaneous symmetry breaking takes place
\cite{Cop,Dvasha}. For D-term inflation \cite{Dinflation} and the
simplest F-term model, the symmetry which is broken at the end of
inflation is a $U(1)$ symmetry and cosmic strings form \cite{prd}
according to the Kibble mechanism \cite{Kibble}. When standard hybrid
inflation is embedded in a grand unified theory (GUT), the breaking
taking place at the end of inflation is that of an intermediate
symmetry group which must eventually break down to the standard model
gauge group and which, requiring that inflation solves the monopole
problem, must contain a U(1) factor \cite{new}. The rank of this
intermediate symmetry group is lowered at the end of inflation and,
barring the formation of unwanted defects, cosmic strings generically
form \cite{prd,new,jrs}.

Cosmic strings can be both a blessing and a curse.  A blessing because
the observation of a cosmic string would be another mean to test
hybrid inflation.  A curse because non-observation constrains the
parameters severely.  The cosmic microwave background (CMB) power
spectrum measured by WMAP points to a predominantly adiabatic
perturbation spectrum, as produced in standard inflation
\cite{WMAP}. The existence of the acoustic peaks excludes cosmic
strings as the main source of perturbations, no more than a 10\%
contribution is allowed ~\cite{Pogosian,Fraisse}.  The cosmic strings
formed at the end of hybrid inflation give a sub-dominant contribution
only if the inflaton-Higgs coupling is much smaller than unity.

In standard F-term inflation the inflaton-Higgs coupling is bound to
the range $10^{-6} -10^{-2}$ \cite{cmb,mairi1}. Hidden sector vacuum
energy and/or moduli vacuum expectation values (VEV) destroy the
successful inflationary prediction for the density perturbations,
unless they are sufficiently small: $\langle z \rangle < 10^{-2} \mpl$
and $\langle W_{\rm hid} \rangle < \TeV \mpl^2$. D-term inflation
fares better with hidden sectors VEV --- they are unconstrained ---
but the coupling constraint is worse~\cite{mairi2}. Not only is the
bound tighter (the strings which form at the end of inflation are BPS
and BPS-strings have a larger tension), there is in addition an upper
limit on the gauge coupling: $g \lesssim 10^{-2}$.  And although small
Yukawa couplings are not unheard of in nature (think of the
electron-Yukawa), no such thing can be said for the known gauge
couplings.

In the literature on F-term hybrid inflation often a spectral index
different from unity is claimed $n_s - 1 \approx 1/N_Q \approx -0.02$
with $N_Q = 50-60$ the number of e-folds of inflation; this estimate
goes back to the original GUT Hybrid inflation paper \cite{Dvasha}.
Indeed, this estimate holds for large enough couplings, in the regime
where the inflaton VEV during inflation is much larger than the
critical VEV where some of the Higgs fields become tachyonic.
However, in the small coupling regime allowed by the CMB data, the
inflaton VEV during inflation is of the order of the critical VEV, and
this approximation breaks down~\cite{Kallosh,moroi,LindeT}.  In this
small coupling regime the spectral index close to a scale invariant
Harrison-Zeldovich spectrum with $n_s =1$. This is more than three
sigma away from the latest WMAP data which strongly favors a
red-tilted spectrum $n_{s, WMAP} -1 = 0.95 \pm 0.015 - 0.019$. Whereas
small couplings can be shrugged off as merely an aestethic question,
the absence of a spectral tilt can exclude hybrid inflation in its
standard form if the WMAP results hold. We note however that for an
appreciable tensor contribution $r$ the best fit value for $n_s =0.98$
\cite{WMAP}.  Although the tensor perturbations from hybrid inflation
are negligibly small, the cosmic string induced vector and tensor
perturbations can be quite high~\cite{Seljak}. This can in principle
be tested using B-type polarization of the CMB.

In this paper we take a closer look at the parameter space for hybrid
inflation. We discuss changes to standard hybrid inflation that can
improve or even alleviate the CMB constraints due to the existence of
cosmic strings.  There are three different strategies:

\begin{itemize}
\item{Lower the scale of symmetry breaking (SSB)}
\item{The strings formed at the end of inflation are unstable}
\item{No strings form at the end of inflation}
\end{itemize}

This paper is organized as follows.  In the next section we review
standard F- and D-term inflation, and the bounds from cosmic string
production. We discuss the one-loop contribution to the potential, and
argue that in realistic GUTs it is smaller than previously thought.
In section 3 we discuss the various possibilities to lower the
symmetry breaking scale: adding terms to the super- or K\"ahler
potential, adding couplings to other fields as in warm inflation,
invoking the curvaton or inhomogeneous reheat scenario, and
entertaining the possibility of varying constants. In Section 4 we
show that the strings which form at the end of inflation are not
necessarily stable; unstable strings do not contribute to CMB
anisotropies. They may still contribute to the baryon asymmetry of the
universe \cite{prl,lepto2}. In section 5 we review scenarios in which
the gauge symmetry is already broken during inflation, and hence no
defects form. Examples are smooth and shifted inflation.
%We discuss
% models in which cosmic strings may form at a lower energy scale.  
We conclude in section 6.

\section{Standard hybrid inflation}

In this section we review standard F- and D-term inflation.  Cosmic
strings form at the end of inflation.  CMB data constrain the model
parameters.

\subsection{F-term inflation}

The superpotential for standard hybrid inflation is 
\cite{Cop,Dvasha}
\be 
W_{\rm inf} = \kappa S( \bar{\Phi} \Phi - M^2),
\label{W}
\ee
with $S$ a gauge singlet superfield, the slowly rolling inflaton field, and
$\Phi$, $\bar{\Phi}$ Higgs superfields in $\mcN$-dimensional complex
conjugate representations of a gauge group $G$.  A suitable U(1)
R-symmetry under which $W$ and $S$ transform in the same way ensures
the uniqueness of this superpotential at the renormalizable level.
The symmetry breaking scale $M$ and the singlet-Higgs coupling
$\kappa$ can be taken real without loss of generality.

Assuming chaotic initial conditions the fields get trapped in the
inflationary valley of local minima at $|S| > S_c = M$ and $|\Phi| =
|\bar{\Phi}| = 0$. The potential is dominated by a constant term $V_0
= \kappa^2 M^4$ which drives inflation.  Inflation ends when the
inflaton drops below its critical value $S_c$ (or when the second
slow-roll parameter $M$ equals unity, whatever happens first) and the
fields roll toward the global SUSY preserving minima of the potential
$|\Phi| = |\bar{\Phi}| =M$ and $|S|=0$.  During this phase transition
$G$ is broken to some subgroup $H$, and cosmic strings form
\cite{prd,new,jrs}.  Topological strings form if $\pi_1(G/H) \neq I$
and embedded strings form if $\pi_1(G_{\rm emb}/H_{\rm emb}) \neq I$
with $G_{\rm emb} \subseteq G$ and $H_{\rm emb} \subseteq
H$~\footnote{Their stability of embedded strings is a dynamical
question, see section \ref{s:embedded}.}.

The scalar potential including all corrections is \cite{cmb,LindeT}
\bea 
\frac{V}{\kappa^2 M^4} &=& 1 +
\frac{\kappa^2 \mcN_l}{32 \pi^2}
\Big[ 2 \ln\(\frac{\kappa^2 M^2 x^2}{\Lambda^2}\) + (x^2+1)^2
\ln(1+x^{-2}) + (x^2-1)^2 \ln (1-x^{-2}) \Big] 
\nonumber \\
&+& 
2 x^4 \( \frac{M}{\mpl}\)^4 + |a|^2 x^2 \( \frac{M}{\mpl}\)^2
+ A \frac{m_{3/2}}{M} x,
\label{V_F}
\eea
with $\mpl = (8\pi G)^{-1/2} \simeq 2.44 \times 10^{18} \GeV$ the
reduced Planck mass, $\mcN_l$ the number of fields contributing to the
1-loop correction, and $\Lambda$ a cutoff scale. We used the notation
$x = |S|/M$ so that $x\to 1$ at the critical point. The first term
gives the tree-level superpotential contribution. The logarithmic
terms are the 1-loop radiative corrections, given by the
Coleman-Weinberg potential (see Eq.~(\ref{CW}) below). We have assumed
a minimal K\"ahler potential to calculate the supergravity correction
--- the terms on the second line of Eq.~(\ref{V_F}). The possible
hidden sector is parameterized as $\langle z \rangle = a \mpl, \quad
\langle W_{\rm hid}(z) \rangle = m_{3/2} \e^{-\frac12 |a|^2} \mpl^2$,
with $m_{3/2}$ the gravitino mass; the cosmological constant in the
global minimum is set to zero by hand.  Further we defined $A= 4
\cos(\arg m_{3/2} - \arg S)$, where we have assumed that $\arg S$ is
constant during inflation. All sub-dominant terms are dropped.  More
details concerning the potential can be found in \cite{cmb}.

\subsection{Coleman-Weinberg potential}

The scalar potential of SUSY hybrid inflation is flat at tree
level. The flatness of the potential is lifted by the loop corrections
which are non-zero during inflation because SUSY is broken
\cite{Dvasha}, and by SUGRA corrections \cite{cmb}. In the major part
of the parameter space loop corrections dominate. In this section we
calculate the loop corrections in the case of SUSY GUT hybrid
inflation and we show that the dimension of the Higgs fields does not
enter in the formula for any realistic model, as opposed to what it is
found in the literature.

SUSY is spontaneously broken along the inflationary valley by a non
zero $F_S$-term. This translates into a mass splitting between boson
and fermion mass eigenstates in the Higgs multiplets. If the dimension
of the Higgs representation $\mcN=1$ the mass spectrum is the
following: There are two complex bosons with $S$ dependent masses
\begin{equation}
m^2_\pm =  (x^2 \pm 1) (\kappa M)^2 
\label{splitb}
\end{equation}
and two Weyl fermions with $S$ dependent masses
\begin{equation}
\tilde{m}^2_\pm =  x^2 (\kappa M)^2.
\label{splitf}
\end{equation}
If the representation of the Higgs fields is $\mcN$-dimensional, there
are $\mcN$ such split double pairs. As $x \to 1$, $\mcN$ of the Higgs
bosons becomes massless, which triggers the phase transition that ends
inflation. 

The mass spectrum can be calculated during the entire inflationary
period, and the loop corrections are continuous \cite{susyb}. They are
given by the Coleman-Weinberg formula \cite{CW}:
\be
V_{\rm CW} = \frac{1}{64 \pi^2}
\sum_i(-)^{F_i}\ m_i^4 \ln \frac{m_i^2}{\Lambda^2}.
\label{CW}
\ee
where $(-)^{F_i}$ shows that bosons and fermions make opposite
contributions. With $\mcN$ mass splitted double pairs
Eqs.~(\ref{splitb}, (\ref{splitf}), we get
\begin{equation}
\frac{V_{CM}}{\kappa^2 M^4} = 
\frac{\kappa^2 \mcN_l}{32 \pi^2}
\Big[ 2 \ln\(\frac{\kappa^2 M^2 x^2}{\Lambda^2}\) + (x^2+1)^2
\ln(1+x^{-2}) + (x^2-1)^2 \ln (1-x^{-2}) \Big] \label{loop}
\end{equation}
where $\mcN_l = \mcN$ has been used in the literature. $\mcN$ can be
very large, and the larger $\mcN_l$, the smaller the parameter space
for inflation allowed by the CMB data. For example in the case of
$SO(10)$ GUT, $\mcN = 16$ or $\mcN = 126$. 

In a realistic model, however, we need more than the $\Phi$ and
$\bar{\Phi}$ Higgs fields to do the full symmetry breaking of the GUT
gauge group $G_{\rm GUT}$ down to the standard model gauge group
$G_{\rm SM}$. The Higgses $\Phi$ and $\bar{\Phi}$ lower the rank of
$G_{\rm GUT}$ upon acquiring a VEV, i.e., they spontaneously break a
diagonal generator of $G_{\rm GUT}$, and they possibly break other
non-diagonal generators of $G_{\rm GUT}$. But in general they cannot
do the full breaking of $G_{\rm GUT}$ down to $G_{\rm
SM}$\footnote{The only GUT which strictly needs only the $\Phi$ and
$\bar{\Phi}$ fields to do the symmetry breaking of $G_{GUT}$ down to
$G_{SM}$ is Pati-Salam $SU(4)_c \times SU(2)_L \times SU(2)_R$ with
$\Phi$ and $\bar{\Phi}$ in the $(4,2,2)$ and $(\bar{4},2,2)$
representation. However even in this case one must introduce extra GUT
Higgs fields to give superheavy mass to some components of $\Phi$ and
$\bar{\Phi}$ which mediate rapid proton decay \cite{leontaris}.}.  Any
GUT model has to satisfy two additional constraints.  Firstly, it must
predict unification of the gauge coupling constants and hence there
cannot be two many lights Higgs components which would spoil
this. Secondly, some components of the Higgs multiplets contain color
triplets which can mediate rapid proton decay, and they must acquire
superheavy masses via couplings to other GUT Higgs fields. 

Now the symmetry breaking taking place at the end of standard hybrid
inflation cannot be the breaking of $G_{\rm GUT}$ but has to be the
breaking of some subgroup $H \subset G_{\rm GUT}$ such that monopoles
form before inflation.  The GUT Higgs fields which break $G_{\rm GUT}$
down to $H$ before inflation acquire GUT scale VEVs and can couple
with $\Phi$ and $\bar{\Phi}$.  This leads to superheavy (GUT scale)
masses for $\mcN - \mcN_l$ components of $\Phi$ and $\bar{\Phi}$. The
$\mcN_l$ fields which remain light are the components of the fields
contained in $\Phi$ and $\bar{\Phi}$ which break $H$.  The heavy
fields decouple and $\mcN$ appearing in Eq.~(\ref{loop}) should be
replaced by the number of light fields $\mcN_l$.  Indeed, the
Coleman-Weinberg potential for $\mcN_l$ light and $\mcN-\mcN_l$ heavy
fields is $V_{CW} = \mcN_l V_{1} (x) + (\mcN-\mcN_l) V_{1}
(x+\frac{m_{\rm GUT}}{\kappa M})$ with
\be
V_{1} (x) = \frac{\kappa^2}{32\pi^2}
\[2\ln\frac{(\kappa M)^2x^2}{\Lambda^2} 
+ (1+x^2)^2 \ln(1+x^{-2}) +(1-x^2)^2 \ln(1-x^{-2})\].
\ee
where $m_{\rm GUT}$ is the mass of the superheavy components of $\Phi$ and
$\bar{\Phi}$. Inflation ends when the slow roll parameter $\eta \propto
V_{CW}''$ exceeds unity, which still happens in the limit $x \to 1$ as
some of the light bosonic fields become tachyonic.  The SSB scale
scales with $V'_{\rm CW}$, see Eqs.~(\ref{dTphi},\ref{estimate}).  For
$m_{\rm GUT} \gg \kappa M$
\be
V'_{CW} \sim \mcN_l V'_1(x) + \frac{\kappa M}{m_{\rm GUT}}  \mcN_h  V'_1(x)
\approx \mcN_l V'_1(x),
\ee
and it is only the number of light fields that enters in the
determination of $M$.

In GUT models which predict massive neutrinos via the see-saw
mechanism, and where $B-L$ strings form at the end of inflation
\cite{prd1,BL}, the symmetry breaking at the end of inflation can be
$H = G_{LR} = SU(3)_c \times SU(2)_L \times SU(2)_R \times U(1)_{B-L}$
or $H = SU(3)_c \times SU(2)_L \times U(1)_R \times
U(1)_{B-L}$\footnote{In these model non-thermal leptogenesis takes
place at the end of inflation; it is a competion between leptogenesis
from reheating and from cosmic string decay
\cite{prl,lepto2,lepto1}.}. If $G_{LR}$ is the intermediate symmetry
group, the only component of the $\Phi$ and $\bar{\Phi}$ fields which
remain massless can either transform as an $SU(2)_R$ doublet, and
$\mcN_l =2$, or as an $SU(2)_R$ triplet, and $\mcN_l=3$. We give an
explicit example in the appendix how this is achieved. In the case of
SO(10), $\mcN_l = 2$ when $\mcN = 16$ and $\mcN_l = 3$ when $\mcN
=126$.

\subsection{D-term inflation}

In D-term inflation the vacuum energy driving inflation originates
from a D-term, to wit from a non-zero Fayet-Illiopoulos (FI) term
$\xi$~\cite{Dinflation}.  This is only possible for an Abelian theory.
The superpotential and D-term potential are
\bea
W &=& \kappa S \bar{\Phi} \Phi \\
V_D &=& \frac{g^2}{2} (\xi + |\bar{\Phi}|^2 - |\Phi|^2)^2
\eea
The Higgs fields have opposite charges which we normalize to
unity. Inflation takes place as the uncharged field $|S| > S_c =
(g/\kappa) \sqrt{\xi}$, and $|\Phi|=|\bar{\Phi}| =0$, and is driven by
the vacuum energy $V_0 = g^2 \xi^2 /2$.  Inflation ends with a $U(1)$
breaking phase transition, in which the field approach their vacuum
values $|S|=|\bar{\Phi}|=0$, $|\Phi| = \sqrt{\xi}$.  The strings which
form at the end of D-inflation are BPS states.  The theory can be
coupled to gravity~\footnote{The $U(1)$ that leaves invariant the FI
term is a combination of the flat space $U(1)$ and an R-symmetry.  The
consequent shift in charges gives only sub-dominant contribution to
the potentials.  We also note that extra singlets are needed for the
theory to be anomaly free \cite{FISUGRA}.}.  Assuming minimal K\"ahler
and gauge kinetic function, the potential is
\be
\frac{V}{g^2\xi^2/2} = 1+
\frac{g^2}{16 \pi^2}
\Big[ 2 \ln\(\frac{g^2 \xi x^2}{\Lambda^2}\) + (x^2+1)^2
\ln(1+x^{-2}) + (x^2-1)^2 \ln (1-x^{-2}) \Big]
\label{V_D}
\ee
where we introduced the notation D), with $z$ a canonically
normalized moduli field.  This will be di
\be
x = \frac{\kappa |S|}{g \sqrt{\xi}} \e^{|S|^2/(2\mpl^2)}.
\label{x_D}
\ee
The potential is a sum of the tree-level D-term potential (first
term), and the radiatively induced Coleman-Weinberg potential (the
logarithmic terms).  The mass splitting between fermions and bosons
responsible for the CW-potential is the same as in F-inflation
Eqs.~(\ref{splitb},\ref{splitf}) after setting $\kappa M \to g
\sqrt{\xi}$ and $\mcN \to 1$. The SUGRA correction is the exponential
factor in the definition of $x$, which enters the CW-potential via the
exponential corrections to the mass eigenstates~\footnote{These
corrections can be neglected in the F-term case where $|S| \ll \mpl$
during inflation.}~\footnote{In D3/D7 inflation, which is a particular
stringy version of D-term inflation, no exponential corrections will
appear in the mass splitting~\cite{Kallosh}.}.  Since $W_{\rm infl} =
0$ during inflation, a non-zero $\langle W_{\rm hid} \rangle$ does not
influence inflation.  A non-zero moduli VEV with minimal K\"ahler can
be incorporated by replacing $|S|^2 \to |S|^2 + |z|^2 $ in the
exponential in Eq.~(\ref{x_D}), with $z$ a canonically normalized
moduli field.  This will be discussed in more detail in
section~\ref{s:kahler}.

\subsection{Density perturbations}

The contributions of cosmic strings and of the inflaton to the
quadrupole temperature anisotropy add up independently \cite{prd}
\be
\( \frac{\delta T }{T} \) =
\sqrt{
\( \frac{\delta T }{T} \)_{\rm inf}^2
+ \( \frac{\delta T }{T} \)_{\rm cs}^2
},
\label{dTtot}
\ee
and should match the observed value $(\delta T/T) = 6.6 \times
10^{-6}$ \cite{WMAP} \footnote{The tensor perturbations $(\delta
T/T)_{\rm tens} \sim 10^{-2} H/\mpl$ are small, and are neglected}.

The string induced perturbations are proportional to the string
tension $(\delta T/T)_{\rm cs} = y G \mu$, with $\mu$ the tension and $y$
parameterizes the density of the string network. Recent simulations
predicts $y=9 \pm 2.5$ \cite{Landriau}, but values in the range
$y=3-12$ can be found in the literature~\cite{Allen,approx}.  The
string tension is~\cite{Hill}
\be 
\mu = 2 \pi M^2 \theta(\beta),
\quad  {\rm with} \quad
\theta(\beta) = \left \{
\begin{array}{lll}
1 & \;\; D{\rm -term} \\
\sim 2.4 \ln (2/\beta)^{-1}  & \;\; F{\rm -term}
\end{array}
\right.
\label{mu}
\ee
The function $\theta$ encodes the correction away from the BPS limit,
and only applies to F-strings. Here $\beta = (m_\phi / m_A)^2 \simeq
(\kappa/g_{\rm GUT})^2$ with GUT coupling $g_{\rm GUT}^2 \approx
4\pi/25$.  Requiring the non-adiabatic string contribution to the
quadrupole to be less than 10\% gives the bound \cite{cmb}
\be
G \mu < 2.3 \times 10^{-7} \( \frac{9}{y} \) 
\quad \Rightarrow \quad
M < 2.3 \times 10^{15} 
\sqrt{\frac{(9/y)}{\theta(\beta)}}.
\label{Pogosian}
\ee

The inflaton contribution to the quadrapole is
\be
\( \frac{\delta T }{T} \)_{\rm inf}
= \frac{1}{12\sqrt{5} \pi \mpl^3} \frac{V^{3/2}}{V'} 
\Bigg|_{\sigma = \sigma_Q},
\label{dTphi}
\ee
with a prime denoting the derivative w.r.t. the real normalized
inflaton field $\sigma = \sqrt{2} |S|$, and the subscript $Q$ denoting
the time observable scales leave the horizon, which happens $N_Q
\approx 60$ $e$-folds before the end of inflation.  The number of
e-foldings before the end of inflation is
\be N_Q = \int_{\sigma_{\rm end}}^{\sigma_Q} \frac{1}{\mpl^2}
\frac{V}{V'} \dd \sigma
\label{N}
\ee
with $\sigma_{\rm end}$ the inflaton VEV when inflation ends, which is
the value for which the slow roll parameter $\eta = \mpl^2
({V''}/{V})$ becomes unity or the Higgs field mass becomes tachyonic,
whichever happens first~\footnote{The first slow roll parameter
$\epsilon = \frac12 \mpl^2 ({V'}/{V})^2 \ll \eta$ can be neglected in
hybrid inflation.}. The spectral index is $n_s-1 \approx 2\eta$.
Finally, we note that the Hubble constant during inflation is $H^2 =
\frac{V}{3\mpl^2}$.

\subsection{Results}

\begin{figure}
\begin{center}
\leavevmode\epsfysize=5.5cm \epsfbox{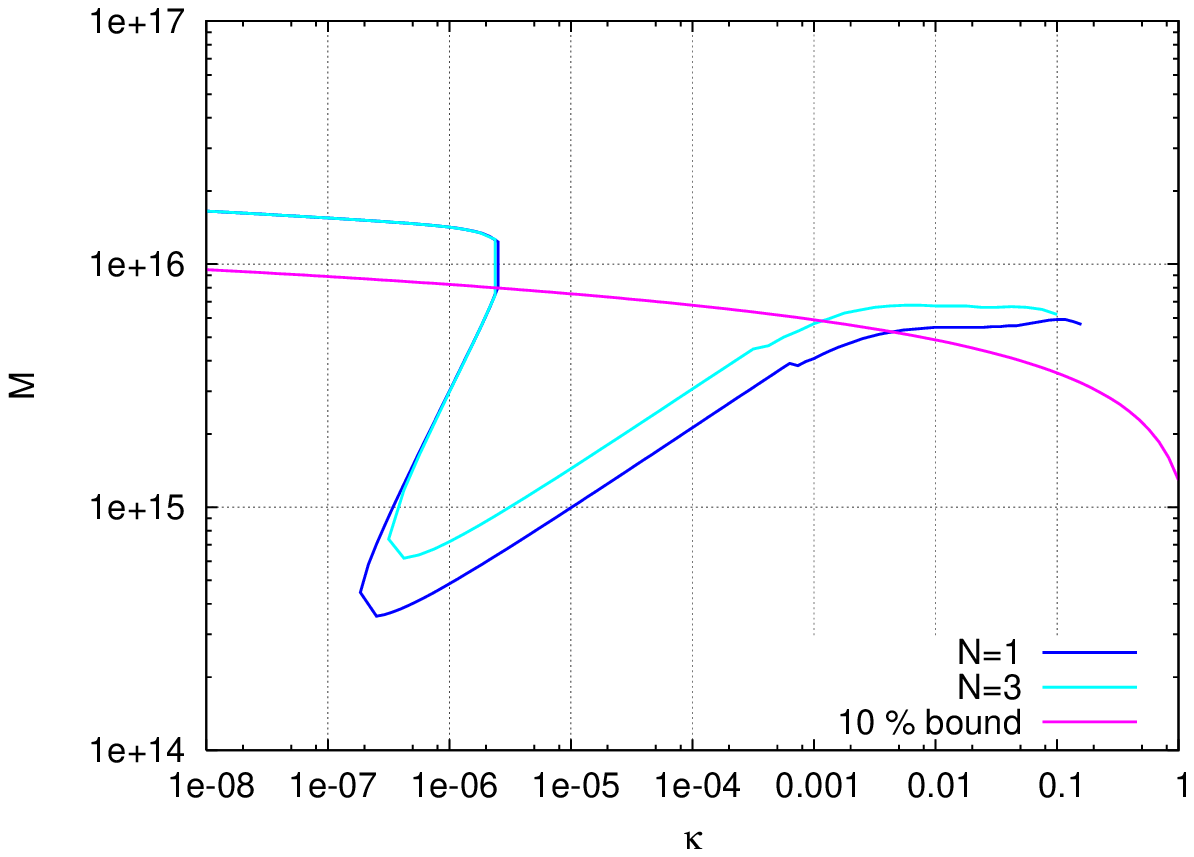}
\leavevmode\epsfysize=5.5cm \epsfbox{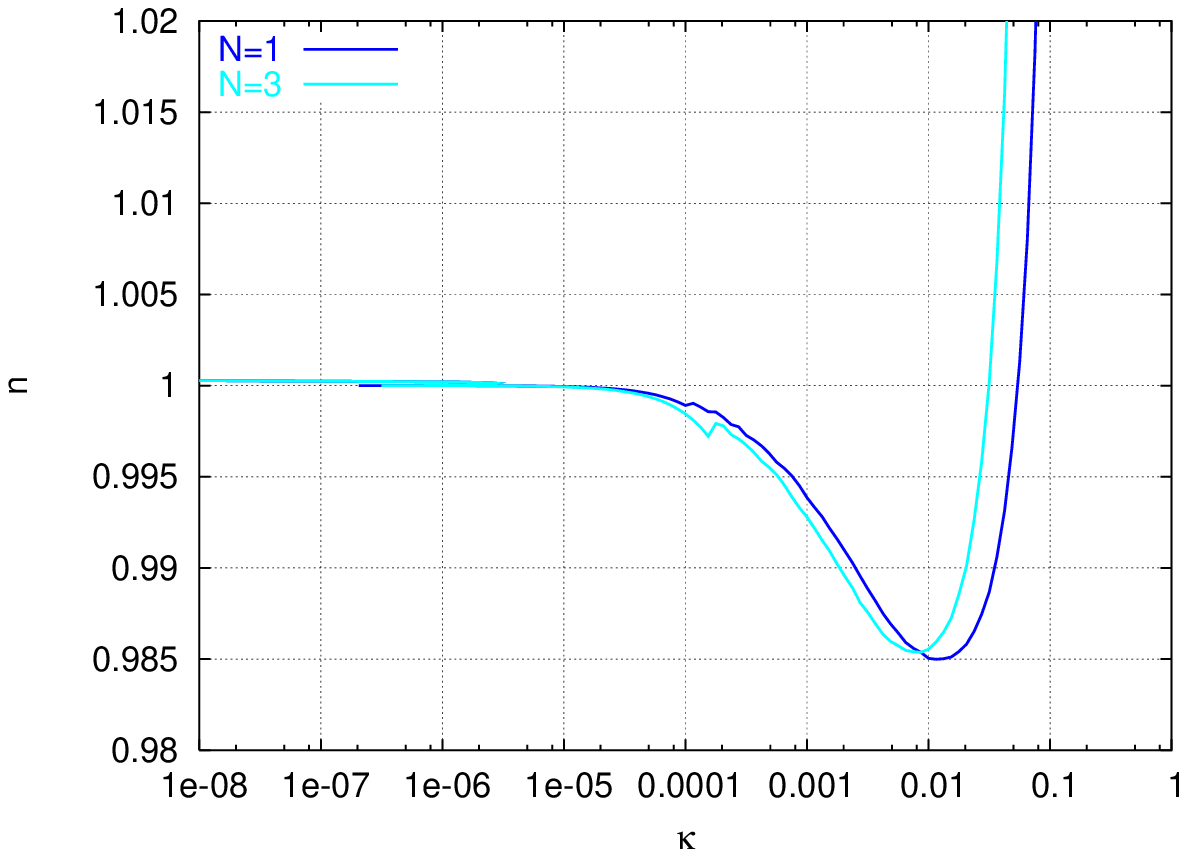}
\caption{$M$ vs. $\kappa$ and $n_s$ vs. $\kappa$ for $\mcN=1,3$ 
in F-term inflation.}
\label{F:F}
\end{center}
\end{figure}

We start with a discussion of $F$-term inflation; more details can be
found in Ref.~\cite{cmb}. Normalizing the temperature anisotropy to
the observed value gives the symmetry breaking scale $M$ as a function
of the inflaton-Higgs coupling $\kappa$. From Eq.~(\ref{dTphi}) it
follows
\be
M^3 
%= 4 \times 10^{-4} \mpl^3 \kappa^{-3} \tilde{V}' 
\propto \kappa^{-3} \frac{V'}{M^3} .
\label{estimate}
\ee
For couplings $\kappa \ll 1$, $x_Q \approx x_{\rm end} \to 1$ and
$V'_{CW}/M^3 \propto \mcN \kappa^4$.  Hence, in the region where
CW-potential dominates (putting back all constants of proportionality)
$M \approx 2 \times 10^{-2} (\mcN \kappa)^{1/3}$.  When other terms in
the potential dominate over the loop potential $V' > V'_{CW}$, the
scale $M$ increases accordingly and cosmic string bounds are stronger.
The result for the potential Eq.~(\ref{V_F}) with $|a|,m_{3/2} =0$ (no
hidden sector VEV) is shown in Fig.~\ref{F:F}.
At large and very small couplings the non-renormalizable term in $V$
dominates the perturbations ($V'_{NR} > V'_{CW}$), at intermediate
couplings the result is determined by $V'_{CW}$ and we can recognize
the $M \propto \kappa^{-1/3}$ behavior.  No solution exist at large
coupling as the potential is too steep for 60 e-folds to occur.

In the literature one can often read the claim that hybrid inflation
predicts a red-tilted spectral index $n_s -1 \approx 1/N_Q \approx
0.98$~\cite{Dvasha,senoguz}.  This is certainly true for moderately
large couplings, but not for the small couplings comprising most of
the parameter space where the cosmic string contribution is
small. Also for large couplings, large
deviations from $1$ are expected \cite{LindeT}. To clarify this issue
we will discuss the spectral index in some detail.  One can
distinguish three different regimes: 1) large $\mcO(1)$ couplings
where the potential is dominated by the non-renormalizable terms. 2)
moderately large couplings $\kappa \sim 10^{-3} -10^{-2}$ where the
potential is dominated by the Coleman-Weinberg potential and inflation
takes place at inflaton values $x_Q \gg 1$. 3) small couplings $\kappa
\lesssim 10^{-3}$ and inflation takes place at inflaton VEV close to
the critical value $x_Q \approx x_{\rm end} \approx 1$.

At large, order one, couplings the inflaton density perturbations are
dominated by the non-renormalizable term in the potential
Eq.~(\ref{V_F}). The slow roll parameter gets a contribution from both
potential terms: $\eta = \eta_{_{\rm NR}} + \eta_{_{\rm CW}}$.  For
super-Planckian inflaton VEV $\eta_{\rm NR}$ exceeds unity implying
that the potential is too steep for inflation to take place. Towards
the end of inflation the CW-potential starts dominating, and inflation
ends when $\eta_{_{\rm CW}} \approx 1$.  Computing $x_{\rm end}$ from
this and subsequently $x_Q$ from Eq.~(\ref{N}) (in the approximation
that except towards the very end of inflaton the potential is
dominated by the non-renormalizable term), allows to calculate
$\eta(x_Q)$ and thus the spectral index
\be
n_s - 1 \approx \frac{3\mcN\kappa^2}{4\pi^2-2\mcN N_Q \kappa^2}
\hspace{2cm}
{\rm for} \;\kappa^2 < \frac{4\pi^2}{2\mcN N_Q}
\ee
For $\kappa^2 < \frac{4\pi^2}{2\mcN N_Q}\approx 0.3/\mcN$ the flat
part of the potential is too short for $N_Q$ e-folds of inflation.
The spectrum is blue-tilted and diverges in the limit where the
potential just allows for $N_Q$ e-folds.

At smaller couplings the potential during inflation is dominated by
the CW-term.  Expanding $V_{\rm CW}$ for large $x\gg1$ we find $x_Q
\approx \sqrt{\mcN N_Q} \kappa \mpl/(2\sqrt{2} \pi M) \approx
\sqrt{N_Q} x_{\rm end}$.  Indeed $x_Q \gg 1$ and the large $x$
expansion is valid for $\kappa \gtrsim M/\mpl \gtrsim
10^{-2}-10^{-3}$. The spectral index then reads
\be
n_s - 1 \approx \frac{1}{N_Q} \approx 0.983
\ee
where in the second equality we used $N_Q = 60$. It is red-tilted and
differs significantly from a featureless Harrison-Zeldovich spectrum.

At smaller couplings inflation ends when the inflaton is close to the
critical value and $x_{\rm end} \approx 1$.  We can expand the
CW-potential in the limit $x \to 1$ to find
\be
N_Q = \frac{8 \pi^2 M^2}{\mcN \kappa^2 \mpl^2} (x_Q - x_{\rm end})
\ee
Hence for $(N_Q \mcN \kappa^2 \mpl^2)/(8 \pi^2 M^2) \ll 1$ one has
$x_Q \to x_{\rm end} \approx 1$.  The spectral index is
\be
n_s - 1 \approx \epsilon \log[ N_Q \epsilon],  \qquad {\rm with}\; 
\epsilon = \frac{\mcN \kappa^2 \mpl^2}{8\pi^2 M^2} \ll 1
\ee
The small $x$ approximation is valid for $\epsilon \ll 1$, hence for
small enough coupling $\kappa$.  In this region the spectrum is
red-tilted (the log-factor is negative) but indistinguishable from a
scale invariant spectrum.  At small coupling there is another viable
solution where the potential is dominated by the non-renormalizable
term. Both $x_Q \approx x_{\rm end} \approx 1$ for this solution
giving a blue-tilted, but nearly scale invariant spectral index 
$n_s - 1 \approx \frac{24 M^2}{\mpl^2}$.

The behavior of the spectral index just described is confirmed by our
numerical calculations, see Fig.~\ref{F:F}.

To summarize the results for F-term inflation, couplings in the range
$10^{-6} < \kappa < 10^{-2}$ are allowed.  In the range $10^{-6} <
\kappa < 10^{-4}$ the spectral index is indistinguishable from
unity. This region is thus excluded by WMAP-3 at the three sigma level
\cite{WMAP}.  For non-zero $|a|,m_{3/2}$, the corresponding terms in
the potential become important when $V' > V_{CW}$; the corresponding
increase in $M$ is acceptable for $a<10^{-2}$, $m_{3/2} < \TeV$.\\

%F-inflation

There are some differences between the F-term and D-term case.  First
of all, the string bound is stronger since D-strings are BPS and
$\theta =1$ in Eq.~(\ref{mu}).  Another difference is that there are
no non-renormalizable or other terms in the potential.  At very small
and large coupling it is still the loop potential that determines the
density perturbations.  The final difference is that the potential,
and thus the results in D-term inflation, depend both on the Higgs
coupling $\kappa$ as on the gauge coupling $g$.

The coupling regime compactible with the CMB bounds is at small
coupling where $x_Q \approx x_{\rm end} \approx
1$~\cite{Kallosh,moroi}. In this regime the inflaton VEV is $|S| \ll
\mpl$ and the SUGRA corrections can be neglected. The FI term scales
as $\sqrt{\xi} = 3 \times 10^{-2} \kappa^{1/3} \mpl$, similar to the
scaling behavior in F-term inflation for small couplings.  At larger
couplings $x_{\rm end},x_Q \gg 1$ (but still sub-Planckian inflaton
VEV) and the scaling behavior breaks down. In this regime the cosmic
string contribution is no longer negligible.

At very small couplings $\kappa$ the inflaton VEV is large $|S| > S_c
\propto \kappa^{-1}$ and the exponential factor in the potential can
no longer be neglected. This results in $\sqrt{\xi}$ increasing with
smaller coupling ($S \approx \mpl$ right at the minimum of
$\sqrt{\xi}(\kappa)$).  For super-Planckian inflaton VEV $S_Q > \mpl$
we expect higher order terms in the K\"ahler potential to be
important: in this regime the low energy effective field theory breaks
down.  Hence, we disregard this part of parameter space.  The coupling
value $\kappa$ at the minimum of $\sqrt{\xi}(\kappa)$ depends on the
gauge coupling $g$.  The smaller the gauge coupling the smaller the
Higgs coupling at the minimum, and consequently the smaller
$\sqrt{\xi}$.  Only for small enough gauge coupling can the string
contribution be made to agree with the CMB data.

The spectral index can be analyzed in the three different regimes: 1)
at large couplings $\kappa$ $x_Q,x_{\rm end} \gg 1$ but $|S| < \mpl$
so that SUGRA corrections can be neglected. 2) At intermediate
couplings where $x_Q,x_{\rm end} \approx 1$ but $|S| < \mpl$ so that
SUGRA corrections can be neglected. 3) At small couplings still
$x_Q,x_{\rm end} \approx 1$ but $|S| \gtrsim \mpl$ and SUGRA
corrections can no longer be neglected. The analysis done using
minimal K\"ahler breaks down.

In the approximation $x_Q,x_{\rm end} \gg 1$ and SUGRA corrections
absent, the spectral index is:
\be
n_s - 1 \approx - \frac{2\kappa}{2 N_Q g + \kappa}
\ee
The spectrum is red tilted, and for small $g$ is much below unity, at
odds with the CMB data.  The large $x$-expansion is valid for $x_{\rm
end} \gg 1$ or $\kappa > 2 \pi\sqrt{2\xi} /\mpl \sim 10^{-2}$; SUGRA
corrections are absent for $g < 2\sqrt{2} \pi$.  At smaller couplings
$g k < 8 \pi^2 M^2/(N_Q \log[4]\mpl^2)$ one has $x_Q,x_{\rm end}
\approx 1$ and the small-$x$ expansion is valid.  In this limit
\be
n_s - 1 \approx \frac{\kappa^2 \mpl^2}{4\pi^2 \xi} 
\log[\epsilon] 
\ee
with $\epsilon$ some small number.  Hence, just like in F-inflation,
the spectral index approaches scale invariance for small $\kappa$.  At
even smaller $\kappa$ the SUGRA corrections should be taken into
account, which complicates the analysis. One finds again $n_s -1
\approx 0$.

%D-inflation
\begin{figure}
\begin{center}
\leavevmode\epsfysize=5.5cm \epsfbox{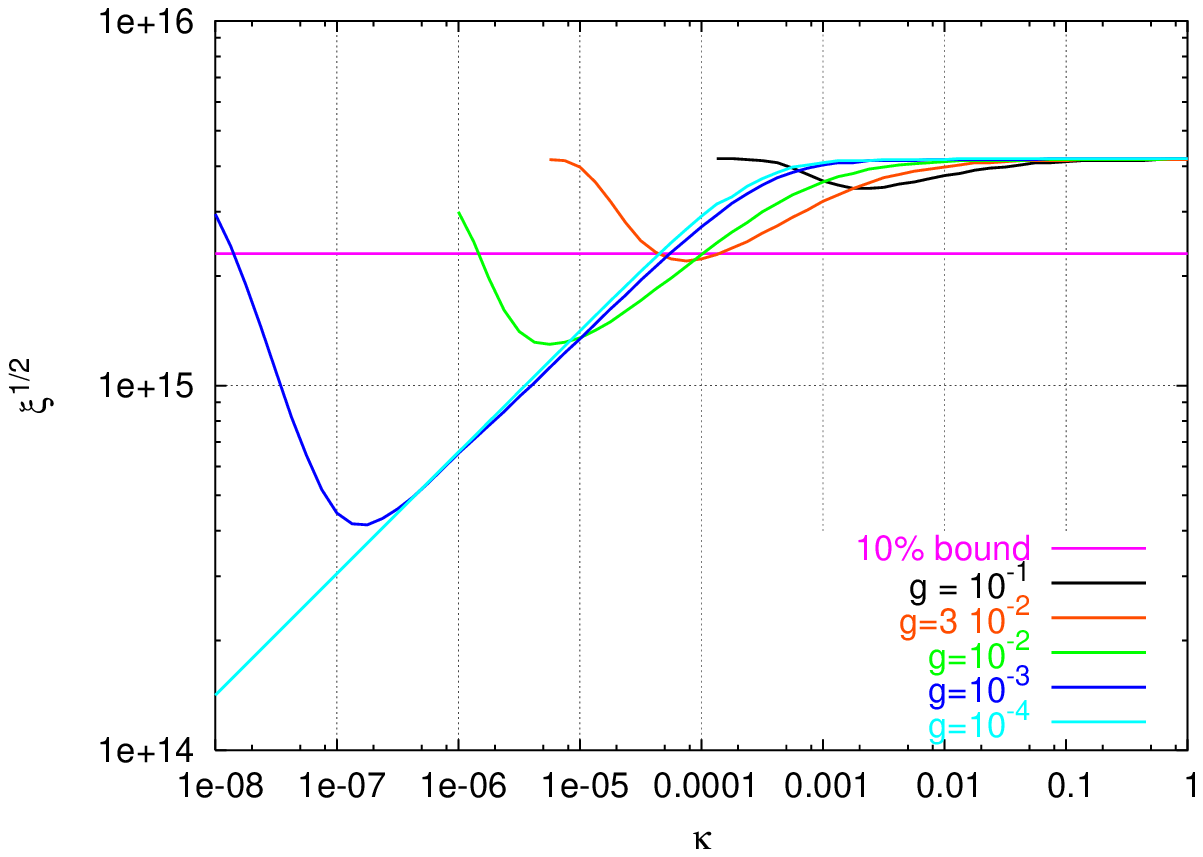}
\leavevmode\epsfysize=5.5cm \epsfbox{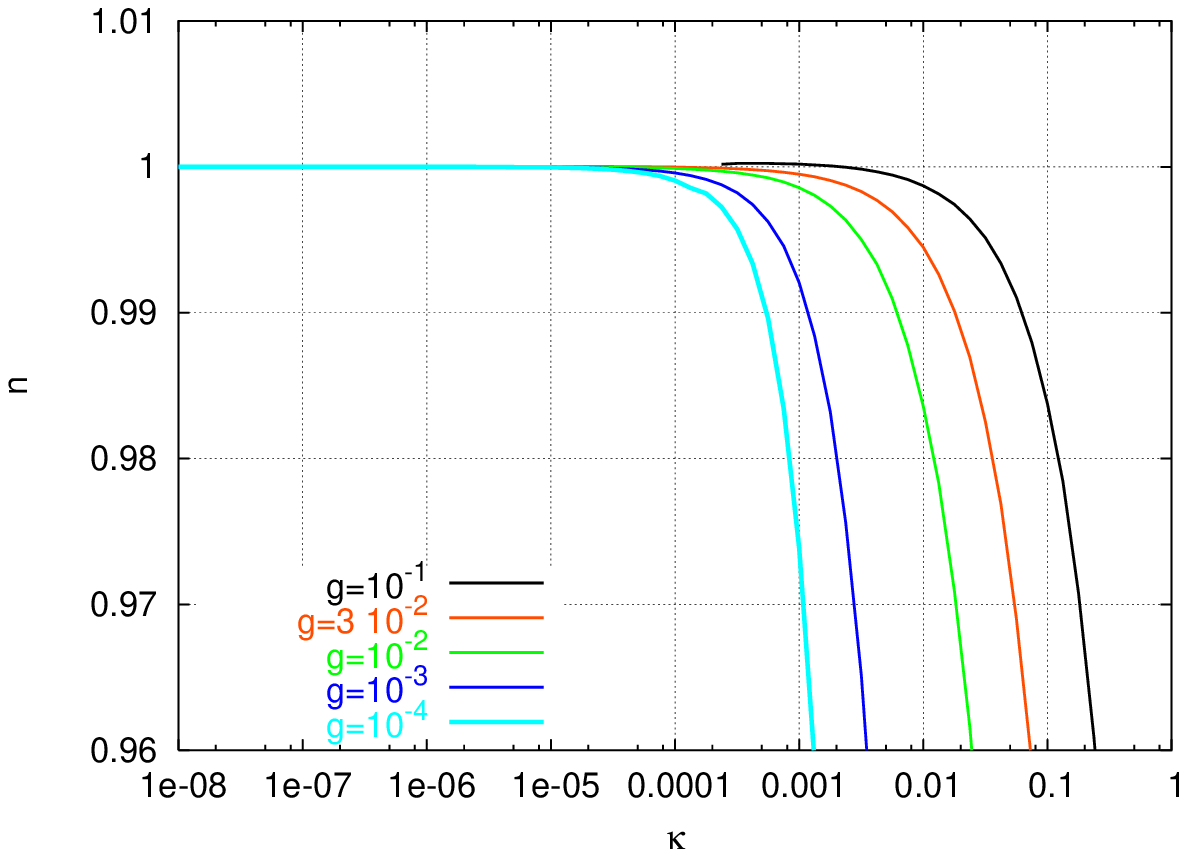}
\caption{$\sqrt{\xi}$ vs. $\kappa$ and $n_s$ vs. 
$\kappa$ for $g=10^{-1}-10^{-4}$ in D-term inflation.}
\label{F:D}
\end{center}
\end{figure}

To summarize the D-term case. The string contribution to the CMB
anisotropies is small for $g < 3\times 10^{-2}$ and $\kappa <
10^{-4}$.  The spectrum is scale invariant at small couplings, and
goes to large negative values at $\kappa > 10^{-2}$, excluded by the
data.  Numerical results are shown in Fig.~\ref{F:D}. We also compared
results for $y=3$ and $y=9$.  Bounds differ by almost an order of
magnitude.  This is the theoretical error in cosmic string
simulations.

\section{Lowering the symmetry breaking scale}

One can try to add additional terms, fields, dynamics to the inflaton
potential, to lower the symmetry breaking scale. In the context of
GUTs, we expect these to be there anyway. In this section we discuss
the various possibilities.  
%As we will see, none of them work satisfactory. 

\subsection{Adding terms to the K\"ahler or superpotential}

It is hard to reduce the symmetry breaking scale by adding extra terms
to the superpotential involving $S$.  As discussed in the last
section, if these extra terms dominated the potential $V'_{\rm extra}>
V'$, the SSB increases, whereas if $V'_{\rm extra}< V'$ they only
affect the density perturbations at a sub-dominant level. To lower the
SSB scale then requires to lower the loop potential $V'_{CW}$, i.e.,
to lower the mass splitting between bosons and fermions.  This is not
easy to achieve. To influence the mass splitting significantly the
extra terms have to be significant during inflation.  This means these
terms generically affect other aspects of inflation as well, and often
the outcome is that they kill inflation rather than revive it.

\paragraph{Mass term for inflaton field.}  

Consider adding a term $W = m_S S^2$ to the superpotential, which
gives a bare mass to the inflaton field.  This term is forbidden by an
R-symmetry in standard hybrid inflation. A scale invariant spectrum
requires $\eta(x_Q) \ll 1$ or $m_S^2 \ll H^2 \sim k^2 M^4/(3\mpl^2)$.
The inflaton mass is necessarily small, and cannot affect the mass
splitting significantly. Explicitly $x^2 \to x(x+m_S/(\kappa M))
\approx x^2 $ in Eqs.~(\ref{splitb},\ref{splitf}).  Similarly for
D-inflation.  Thus, a bare inflaton mass term cannot improve the
bounds, but if too large it kills inflation!

\paragraph{Higher order inflaton coupling}

Replace the inflaton-Higgs coupling in the superpotential term by $W =
\kappa S^n (\bar{\Phi} \Phi)^m$.  Consider first the case $m=1$. It
can be checked that $V'_{CW} \propto n$, and the larger $V'$ the
larger the SSB scale (in this case $M \propto n^{1/3}$). Thus the
standard case with $n=1$ is preferred, not only because it is
renormalizable but also because it gives a lower SSB.  For $m=2$ the
above potential gives rise to smooth inflation in the F-term case, to
be discussed in section \ref{s:smooth}. For $m \geq 1$ in D-term
inflation and $m \geq 2$ in F-term inflation, there is no symmetry
breaking potential and thus no hybrid inflation.

\paragraph{Mass term for Higgs fields.}

Consider another superfield $X$ coupling to the Higgs fields, $W =
 \lambda X \bar{\Phi} \Phi$. This gives a bare mass to the Higgs
 fields.  If $X$ is gauge singlet, assuming $m_X = \lambda \langle X
 \rangle$, the effect it to shift $x \to x + m_X/(\kappa M)$ in
 Eq~(\ref{splitb},\ref{splitf}).  For $m_X > (\kappa M)$ both Higgs
 masses are positive for all values of $x$, and there is no phase
 transition ending inflation.  Since $\epsilon,\eta \ll 1$ for all
 values of $x$, inflation is eternal (driven by a non-zero
 cosmological constant $\kappa^2 M^4$)!  In the opposite limit $m_X <
 \kappa M$ the bare mass is sub-dominant during inflation, and cannot
 change $V'_{CW}$ significantly.

\paragraph{Beyond minimal K\"ahler potential}
\label{s:kahler}
Include the higher order terms in the K\"ahler potential
\be 
K = K(z) + \(1+f_S(|S|^2)\)|S|^2 + \(1+f(|S|^2)\)|\Phi|^2 +
\(1+\bar{f}(|S|^2)\)|\bar{\Phi}|^2 \ee
with $z$ a hidden sector field which has no superpotential coupling to
the inflaton sector.  In F-term inflation with minimal Kh\"aler
potential for all the fields ($f_S,f,\bar{f}=0$), there is no $H^2
|S|^2$ term in the potential due to a miraculous cancellation. One
generally expects higher order terms to be present, which can possible
ruin this cancellation.  Indeed, the first order term $f_S = - a^2/4
|S|^2 + ...$ or a hidden sector term $K(z) = a^2 \mpl^2$
(corresponding to a VEV $\langle z \rangle = a \mpl$ and $z$
canonically normalized) would give a mass term $V = a^2 H^2 |S|^2$.
Hence, inflation can only take place for $a \lesssim 10^{-2}$. Higher
order terms $\mcO(S^4)$ in $K$ are negligible in F-term inflation,
since $S \ll \mpl$.

%D-inflation A & expK
\begin{figure}
\begin{center}
\leavevmode\epsfysize=5.5cm \epsfbox{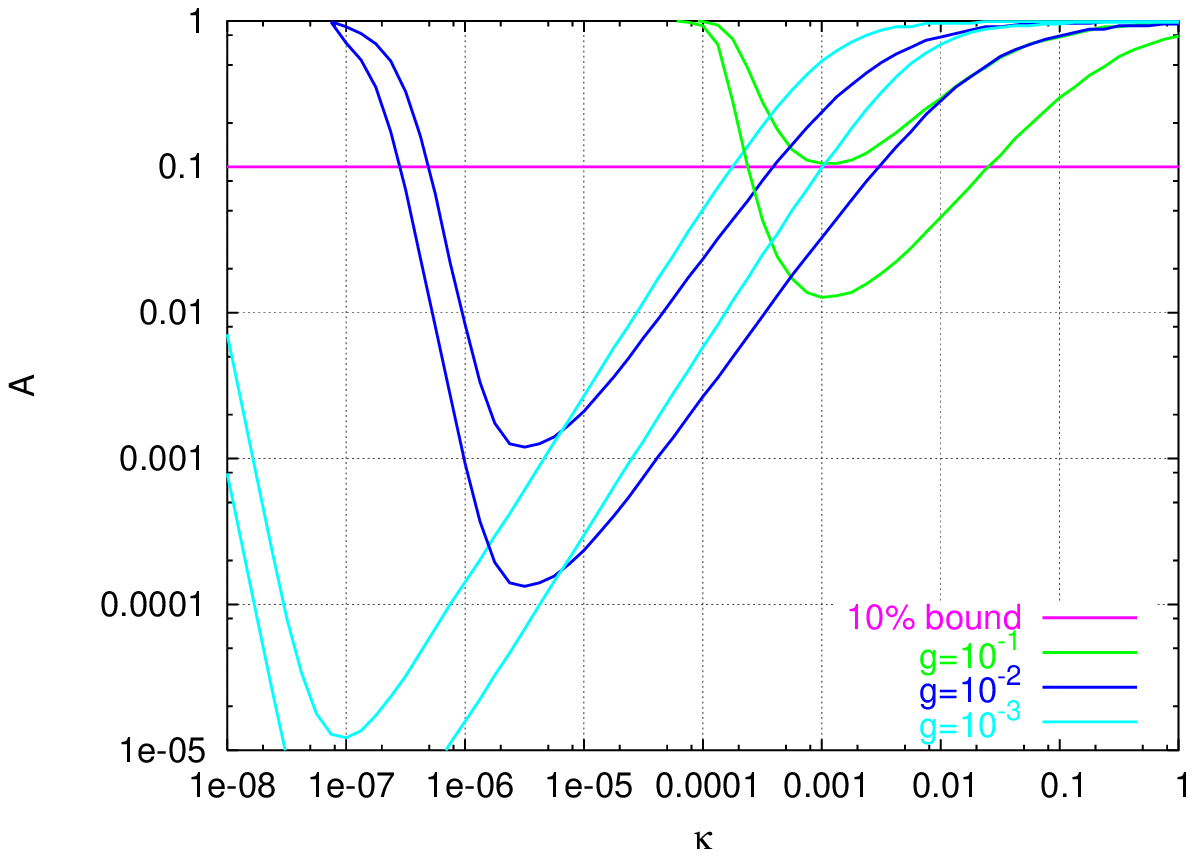}
\leavevmode\epsfysize=5.5cm \epsfbox{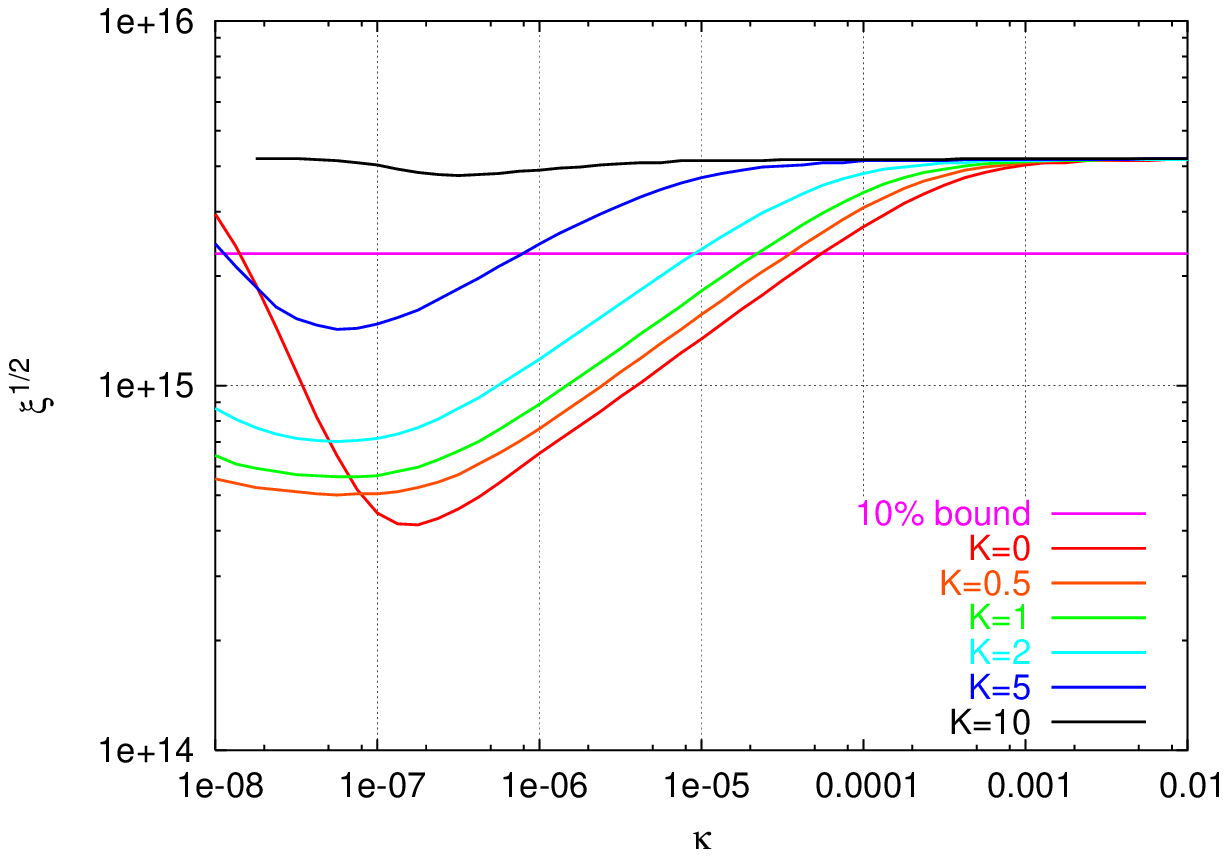}
\caption{Left plot shows $[(\delta T/T)_{\rm cs} / (\delta T/T)_{\rm
WMAP}]^2 $ vs. $\kappa$ for $g=10^{-1},10^{-2},10^{-3}$ and for $y=3$
and $y=9$. Right plot shows $\sqrt{\xi}$ vs. $\kappa$ for $g=
10^{-3}$, $y=9$ and $K=0,0.5,1,5,10 \mpl$.}
\label{F:AK}
\end{center}
\end{figure}

There is no Hubble induced mass term in D-inflation dependent on
$f_S$; non-zero $f_S$ can be absorbed in redefinition of $S$
\cite{Seto}. A non-zero $K(z)$ alters Eq.~(\ref{x_D}) to
\be
x = \frac{\kappa |S|}{g \sqrt{\xi}} \exp\(\frac{|S|^2+K(z)}{2\mpl^2}\).
\ee
The hidden sector becomes important when $K(z)\sim \mpl^2$, which for
$z$ canonically normalized corresponds to $\langle z \rangle \sim
\mpl$.~\footnote{Provided $\partial W/\partial z \neq 0$, otherwise
the term $\exp K$ can be removed from the potential by a rescaling.}
This is shown in Fig.~\ref{F:AK}, where $M(\kappa)$ is plotted for
various values of $K(z)$.  Although the constraint on the modulus is
weaker than in F-inflation where $\langle z \rangle \ll \mpl$ is
needed, this can still be a severe problem for model building, as
moduli are often stabilized at Planckian VEV.

Consider then $K(z)$ and $f_S$ small or absent, and concentrate on the
effects of higher order terms coupling the singlet with the Higgs
fields, parameterized by $f, \bar{f}$.  As shown in Ref.~\cite{Seto}
the main consequence is the replacement
\be
x^2 \to \frac{x^2}{(1+f_+)(1+f_-)}
\ee
in the mass split boson-fermion pairs
Eqs.~(\ref{splitb},\ref{splitf}). Since $|S|_{\rm end} \ll \mpl$, $f,
\bar{f}$ are sub-dominant at the end of inflation, and do not affect
the moment inflation ends.  The inflaton VEV can be large when
observable scales leave the horizon. Remember $M^3 \propto V'_{CW}$
with for $x \gg 1$ \footnote{We write the potential in terms of
$\sigma$ instead of $x$, as this makes explicit whether the inflaton
VEV is below the Planck scale.}
\be
V'_{CW} \propto  \( \frac{2}{\sigma} + \sigma 
-\frac{f'_+}{1+f_+} -\frac{f'_-}{1+f_-}\)
\ee
where we have set temporarily $\mpl = 1$, and prime denotes derivative
w.r.t. $\sigma$. Expanding the $f$-functions $f = c \sigma^2/2 +
{\mathcal O}(\sigma^4)$, we see that the higher order terms alter the
first order results considerably if $2c\sigma/(1+c {\sigma}^2/2) \sim
{\sigma} + 2/\sigma$, which requires both $\sigma \to 1$ and $c>1$.
In this regime the second order term dominates over the first order.
There is no good ground to drop third and higher order terms.  This is
not a satisfactory way to reduce the SSB, as we are entering the
regime where effective field theory breaks down, and the K\"ahler
potential to all orders seems needed.

\subsection{Warm inflation}

%Warm inflation 
\begin{figure}
\begin{center}
\leavevmode\epsfysize=5.5cm \epsfbox{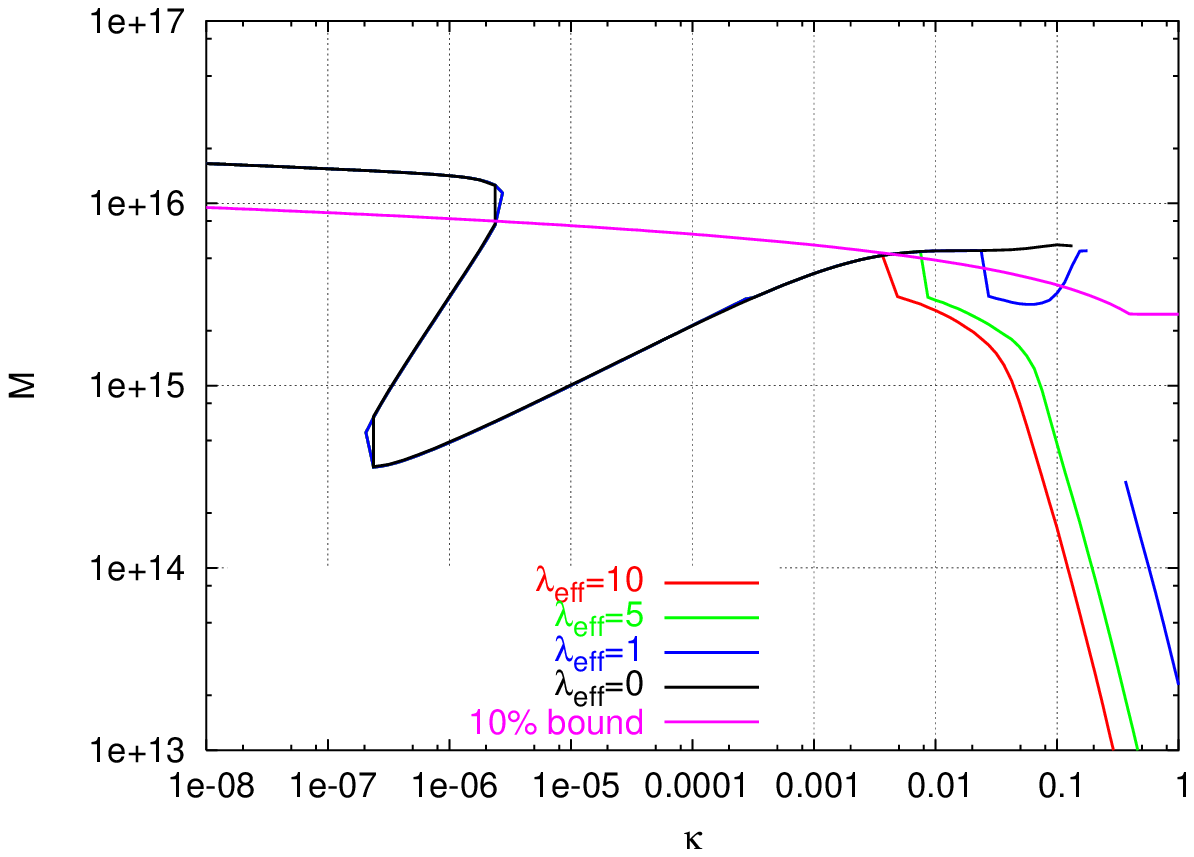}
\leavevmode\epsfysize=5.5cm \epsfbox{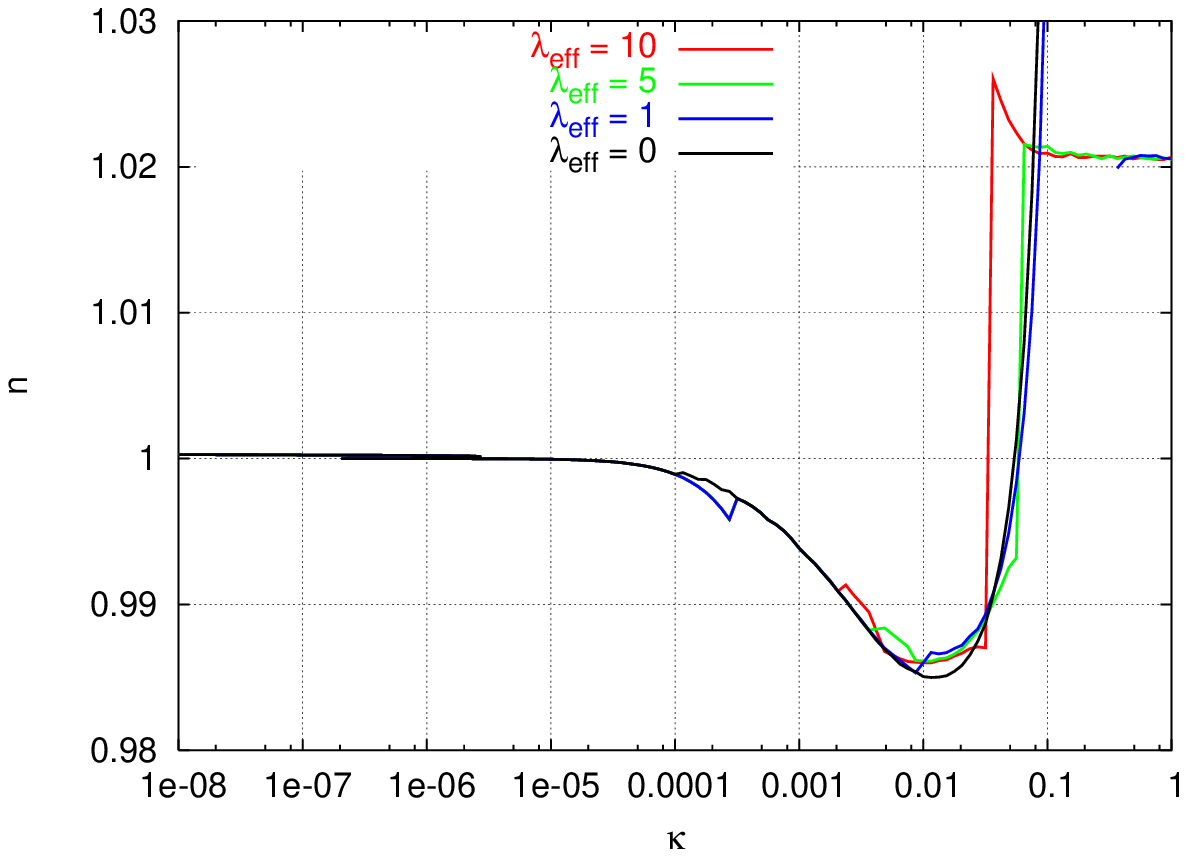}
\caption{$M$ vs. $\kappa$ and $n_s$ vs. $\kappa$ for
$\lambda_{\rm eff}=0,1,5,10$ and $\mcN =1$.}
\label{F:warm}
\end{center}
\end{figure}

If the (indirect) coupling of the inflaton to light fields is large,
it can decay during inflation \cite{berera}. Although the energy
density in radiation remains small throughout inflation, it is
nevertheless possible that $T > H$.  The inflaton fluctuations are
thermal, and enhanced w.r.t. cold inflation.  If in addition the decay
rate is large $\Gamma > H$, the inflaton damping due to decay is
larger than due to the expansion of the universe. This gives an extra
enhancement of the perturbations.  Smaller SSB scale is needed to
obtain the observed quadrupole, and the bounds from cosmic string
production may be avoided.

To implement warm inflation we add a term 
\be
W = \lambda \Phi NN
\label{RHN}
\ee
that allows inflaton decay into light $N$-quanta.  Physically, what
happens is that during inflation the slowly changing inflaton field
can excite the Higgs field $\Phi$, which can decay into massless $N$
fields.  The 1-loop dissipation coefficient scales as $\Gamma \sim
\kappa^4 \Phi^2 \Gamma_{\Phi \to N N} \sim \kappa^3 \lambda^2 \Phi$.
An explicit QFT calculation in the adiabatic-Markovian
limit\footnote{valid for $\frac{\dot{\phi}}{\phi} < H < \Gamma$.}
gives\cite{berera,mar}
\be
\Gamma =  \frac{\kappa^3 \lambda^2 \mcN_l N_f M}{2048 \pi^2} 
\frac{x^2}{\sqrt{1+x^2}}
\ee
with $\mcN_l$ the number of intermediate $\Phi$ quanta, and $N_f$ the
number of final states a Higgs boson can decay into. We absorb these
factors in an effective coupling $\lambda^2_{\rm eff} = \lambda^2 \mcN_l
N_f$; all plots are a function of $\lambda_{\rm eff}$.  The
dissipation coefficient enters as a friction coefficient in the
inflaton equation of motion.  

The equations for the density perturbations can be generalized to
include thermal effects. We will list here the important formulas;
more details can be found in \cite{cmb,mar}.  Slow roll inflation ends
when the slow roll parameters $\eta \approx r$ with $r$ the ratio
$r(x) = \Gamma/(3H)$\footnote{The results are fairly independent of
$x_{\rm end}$; using that inflation is ended when $\epsilon=r$ or
$\beta =r$ instead gives visually indistinguishable result.  The
reason is that the $N_Q$ integral is dominated by $x_Q$ and thus
hardly depends on $x_{\rm end}$.}. The dissipative effects
parameterized by $r(x)$ are maximized in the limit
$\kappa,\lambda_{\rm eff}$ large, as this maximizes the decay rate.
The formulas for the density parameters then generalize as follows.
The number of e-folds is
\be 
N_Q = \int_{x_{\rm end}}^{x_Q} 
\frac{1}{\mpl^2} \frac{V}{V'}(1+r) \dd x
\label{w:N_Q}
\ee
The quadrupole temperature anisotropy can be approximated by
\be
\( \frac{\delta T }{T} \)_{\rm infl} = 
\frac{1}{12\sqrt{5}\pi \mpl^3}\frac{V^{3/2}}{V'} 
\(1+r\) \(1+ \( \frac{3 \pi r}{4} \)^{1/4} \) \(1 + \sqrt{\frac{T}{H}}\) 
\label{w:dTphi}
\ee
with r.h.s. evaluated at $x_Q$. The temperature during inflation
follows from $\rho_\gamma = \pi^2 g_* /(30)T^4$ with $g_* \sim 100$
and
\be
\frac{\rho_\gamma}{H^4} = \frac{9}{2} 
\frac{r \epsilon}{(1+r)^2\kappa^2} 
\(\frac{\mpl}{ M}\)^4
\ee
In the limit $r \to 0$ (and thus also $\Gamma \to 0$, $T \to 0$) all
above formulas reduce to those of standard cold inflation, where
dissipative effects are negligible small.  Finally, we give the
generalization of the spectral index in the dissipative regime
\be
n_s - 1 =
\left\{
\begin{array}{lll}
& -\frac{17}{4} \epsilon + \frac32 \eta - \frac14 \beta,  
&\qquad  {\rm for} \; \Gamma < H < T \\
& (-\frac{9}{4} \epsilon + \frac32 \eta - \frac94 \beta) 
(1+r)^{-1},  
&\qquad  {\rm for} \; H < \Gamma,T 
\end{array}
\right. 
\label{n_warm}
\ee
with $\beta = \mpl^2 (\Gamma'/\Gamma) (V'/V)$.

The result for warm inflation are shown in Fig.~\ref{F:warm}.  The
interesting part is the large coupling regime $\lambda_{\rm eff}
\gtrsim 0.1$ and $\kappa \sim 1$ where $\Gamma > H $ and thermal
effects are important.  The density perturbations are enhanced and
consequently a much smaller SSB scale is needed.  The cosmic string
contribution is negligible small.  However, if we look at the spectral
index in this regime, we see that $n_s \approx 1.02$ is blue-tilted,
and thus strongly disfavored by the latest data.  At smaller couplings
there is no inflationary solution with $N_Q= 60$, just as in cold
inflation.  At these and smaller couplings, the thermal effects are
small, and the solution approaches that of cold inflation.

The plots are for F-term inflation, but similar results expected for
D-term.

\subsection{Curvaton or inhomogeneous reheat scenario}

%Curvaton and varying constants
\begin{figure}
\begin{center}
\leavevmode\epsfysize=5.5cm \epsfbox{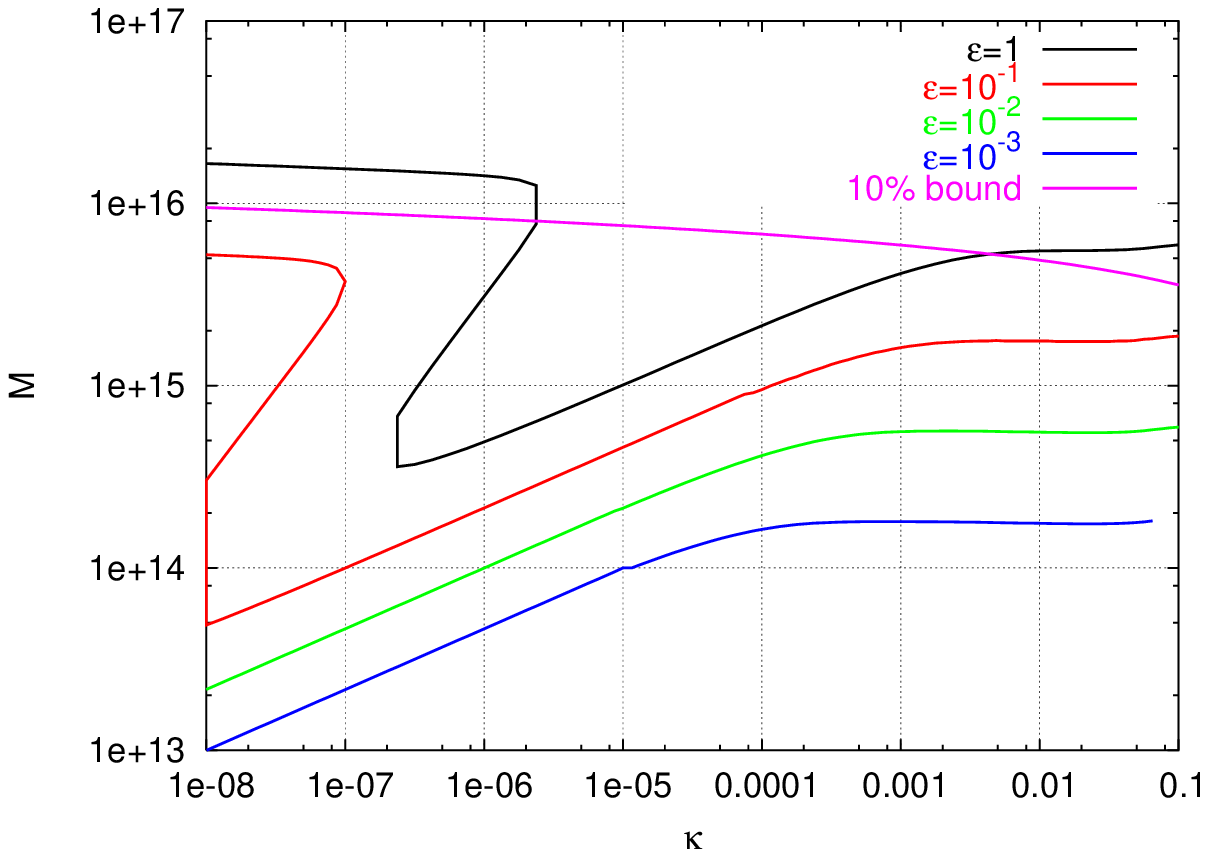}
\leavevmode\epsfysize=5.5cm \epsfbox{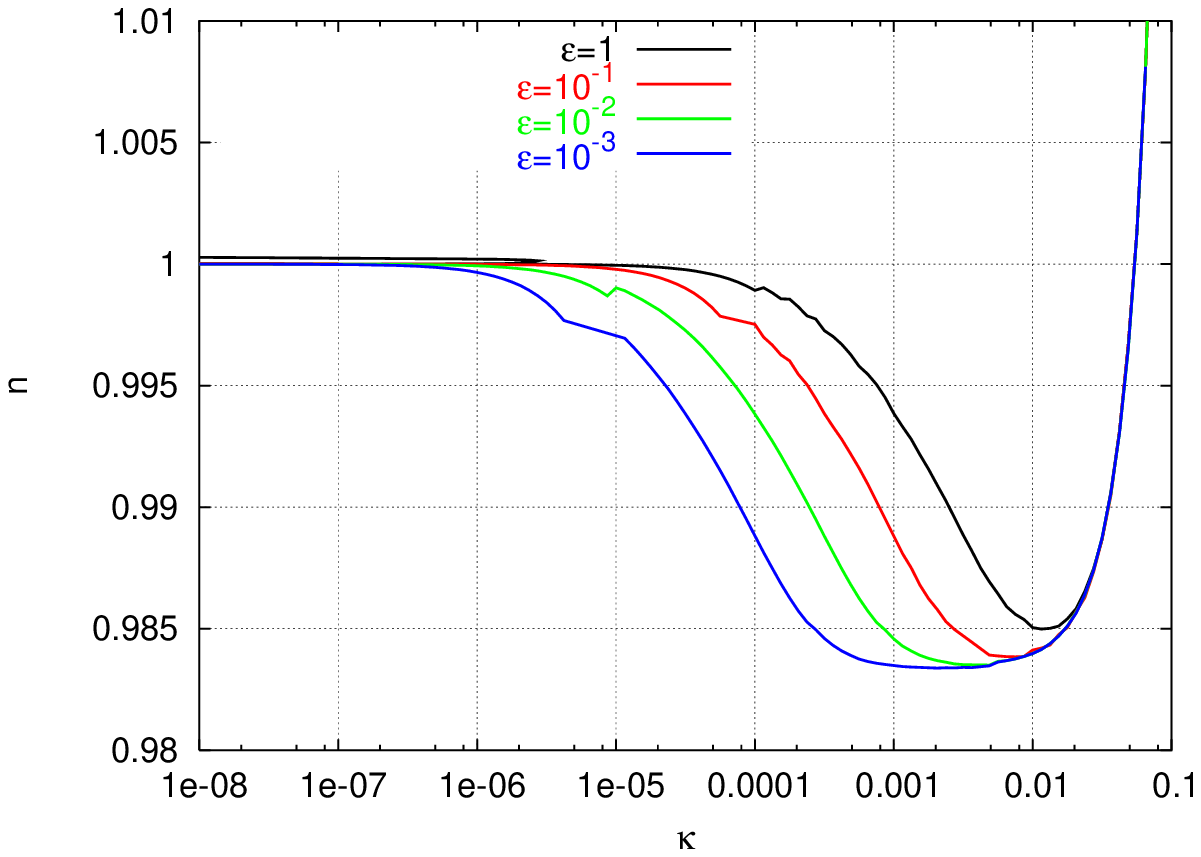}
\caption{$M$ vs. $\kappa$ and $n_s$ vs. $\kappa$ for $(\delta
T/T)_{\rm infl}= \epsilon (\delta T/T)_{_{\rm WMAP}}$ with $\epsilon =
1,10^{-1},10^{-2},10^{-3}$ and $\mcN_l =1$. Spectral index only applies
to the varying constant case and not to curvaton (as there bulk of
perturbations and thus spectral index produced by another field)}
\label{F:curv}
\end{center}
\end{figure}

Since the cosmic string contribution to the temperature anisotropies
is proportional to the inflaton contribution, lowering the latter will
obviously lower the first thereby circumventing all CMB constraints.
If the inflaton contributes only a small fraction to the density
perturbations, $(\delta T/T)_{\rm infl} = \epsilon(\delta T/T)_{_{\rm}
WMAP}$ with $\epsilon < 1$, some other field or sector of the theory
must provide the bulk.  This can be achieved~\cite{mairi1,moroi}
employing the curvaton~\cite{curvaton} or inhomogeneous reheating
scenario~\cite{irs}.

The total temperature anisotropy is given by
\be
\( \frac{\delta T}{T} \) 
= 
\sqrt{
\( \frac{\delta T}{T} \)^2_{\rm infl} + 
\( \frac{\delta T}{T} \)^2_{\rm cs} + 
\( \frac{\delta T}{T} \)^2_{\rm s}
} 
\ee
with $\[ {\delta T}/{T} \]_{\rm s}$ parameterizing the contribution of
the curvaton field or the fluctuating field of the inhomogeneous reheat
scenario.  If this contribution dominates, the inflaton quadrupole is
only a small fraction of the value measured by WMAP:
\be
\( \frac{\delta T}{T} \)_{\rm infl} 
= \epsilon  \( \frac{\delta T}{T} \)_{_{\rm WMAP}}
= \epsilon \, 6.6 \times 10^{-6} 
\label{epsilon}
\ee
This leads, for example, in the intermediate coupling regime where the
CW potential dominates and $x_Q \approx x_{\rm end} \approx 1$ the
inflationary scale $M \propto \epsilon^{1/3}$ is lowered.  Our
numerical calculations show that already for $\epsilon = 0.1$ the
parameter space is opened up to couplings in the range $10^{-7}
\lesssim \kappa \lesssim 10^{-1}$, see Ref.\cite{cmb} for more
details. Results are shown in Fig.~\ref{F:curv}.

We should note that the curvaton scenario in its standard/simplest
set-up can only work for a large enough Hubble parameter during
inflation $H>10^8\GeV$ \cite{lyth,postma}. This requires for example
$\kappa \gtrsim 10^{-4}$ in F-inflation.

\subsection{Varying constants}

The couplings and masses appearing in our inflationary model can
fluctuate in time.  In SUSY theories as well as in string inspired
models, all effective couplings are not constants, but rather
functions of scalar fields in the theory.  Many of these scalar fields
have flat potentials and may only reach their ground state well after
inflation.  There are strong constraints on the variability of
constants at present, but practically none for any variation happening
before the time of nucleosynthesis.

The quantum fluctuations of the inflaton field leave the horizon
during inflation.  The thus produced curvature perturbation remains
constant until horizon re-entry, provided there is no significant
non-adiabatic pressure perturbation. The density perturbations from
inflation are only a function of the model parameters during
inflation.  This statement is based on local conservation of
energy-momentum and is independent of the gravitational field
equations \cite{Wands}.  It holds for any metric theory of gravity,
including scalar-tensor theories or induced four-dimensional gravity.
In such theories, varying constants are general.  For a non-Einstein
version of hybrid inflation see Ref.~\cite{Bento}.

The cosmic string contribution to the CMB originates from strings at
red shifts between the surface of last scattering $z = z_{\rm ls}$ and
the present $z = 0$, though most strings are present at $z = z_{\rm
ls}$ \cite{ShelVil}.  It is clear then that if the parameters in the
theory vary between the time of inflation, when inflaton perturbations
are produced, and last scattering, when cosmic string perturbations
are produced, the proportionality between the two sources of
perturbations changes.  In particular, it may be that the string
contribution is reduced w.r.t. the inflaton contribution.

Consider for example the case that the Planck mass varies between
inflation and present; we write $\mpl(t_Q) = f^{1/3} \mpl(t_0)$.  In
the region of parameter space where the density perturbations are
dominated by the CW potential (and thus $V'$ independent of the Planck
mass) the total temperature anisotropy is
\bea
\( \frac{\delta T}{T} \) 
&=& 
\sqrt{
\( \frac{\delta T(t_Q)}{T(t_Q)} \)^2_{\rm infl} + 
\( \frac{\delta T(t_0)}{T(t_0)} \)^2_{\rm cs}
} \nonumber \\
&=&
\sqrt{
\( {f} \frac{\delta T}{T} \)^2_{{\rm infl},0} + 
\( \frac{\delta T}{T} \)^2_{\rm cs,0}
} 
\approx {f} \(\frac{\delta T}{T} \)_{{\rm infl},0}
\label{varying}
\eea
where $(\delta T /T)_{{\rm infl},0}$ is given by Eq.~(\ref{dTphi})
with all couplings at their present value $t = t_0$, and likewise
$(\delta T /T)_{{\rm cs},0} = y G \mu $ with all couplings at their
present value $t = t_0$. The inflaton contribution is enhanced by a
factor $f$ with respect to the standard situation with a constant
Planck mass $f =1$ and $\mpl (t_Q) = \mpl(t_0)$. 

In general not only one effective coupling can vary but many.  Writing
the inflaton contribution as
\be
\( \frac{\delta T (t_Q)}{T(t_Q)} \)_{\rm infl}  = 
f(\kappa) \( \frac{\delta T}{T} \)_{{\rm infl},0} 
\ee
the inflaton contribution is enhanced by a factor $f(\kappa)$
w.r.t. the standard case.  Note that $f(\kappa)$ can be
$\kappa$-dependent.  The reason is that at different $\kappa$ values
different terms in the potential dominate, and thus the parameter
dependence of the inflaton perturbations depends on $\kappa$.

For $f\neq 1$ and $\kappa$-independent the bounds are the same as in
the curvaton or inhomogeneous reheat scenario with $\epsilon = 1 /f$
the fraction $(\delta T/T)_{{\rm infl},0}$ contributes to the
amplitude measured by WMAP ({\it cf.}
Eqs.~(\ref{epsilon},~\ref{varying} )), see Fig.\ref{F:curv}.

\section{Unstable strings}

If the strings which form at the end of inflation decay before
matter-radiation equality, CMB constraints will be avoided. This is
possible if the strings are embedded, non-topologically stable,
semilocal, or if they nucleate monopole-antimonopoles at a sufficiently
fast rate \cite{embedded}.

\subsection{Embedded strings}
\label{s:embedded}

Cosmic strings form when a group G spontaneously breaks down to a
subgroup H of G if the first homotopy group of the vacuum manifold G/H
is non-trivial $\pi_1(G/H) \neq I$ \cite{Kibble}. It is however
possible to find string solutions even if $\pi_1(G/H) = I$; if
$\pi_1(G_{\rm emb}/H_{\rm emb}) \neq I$ with $G_{\rm emb} \subseteq G$
and $H_{\rm emb} \subseteq H$ embedded strings form
\cite{embedded}. The string stability is then a dynamical question and
it depends on the parameters of the model \cite{ShelVil}. A possible
symmetry breaking at the end of inflation which leads to the formation
of embedded strings and which does not lead to any other unwanted
defect is given by
\begin{equation}
H \times U(1)_X \rightarrow U(1)_Y
\end{equation}
where $H$ a non-abelian and simple group, $U(1)_Y$ is not contained in
$H$ and $U(1)_X \neq U(1)_Y$, i.e. the $Y$ charge generator is a
linear combination of $X$ and a diagonal generator of $H$ which
corresponds to a neutral gauge boson which we call $W_0$. (We give the
same name for the gauge bosons and thee corresponding generators.)
There are two type of strings which form, an abelian string, whose
generator is the abelian generator orthogonal to $Z$ and a non-Abelian
one.  The non Abelian is unphysical \cite{ShelVil}.

Embedded strings are non topological because there is no conserved
topological charge. Their stability is a dynamical question and it
depends on the mixing between $X$ and $W_0$ and on the ratio between
the masses of the Higgs and gauge field forming the string
\cite{ShelVil,embedded,ew}. Stability requires a very large mixing
angle. In the case of $H = SU(2)$ for example, the strings are stable
if $\sin^2(\theta) \gtrsim 0.9$, where $\theta$ is the define by $Y =
\sin(\theta) \, W_0 + \cos(\theta) \, X$ and the gauge field forming
the string $A = \cos(\theta) \, W_0 - \sin(\theta) \, X$
\cite{ew}. For example electroweak strings are unstable \cite{ew}. We
conclude that embedded strings can only be stable if their gauge field
is mostly made out of $X$.

Embedded strings form after hybrid inflation in GUT models which
predict massive neutrino via the see-saw mechanism if $G_{LR}$ is
broken down to the SM gauge group at the end of inflation
\begin{equation}
SU(3)_c \times SU(2)_L \times SU(2)_R \times U(1)_{B-L}\stackrel{\bf
Inflation + Embedded \, Strings}\longrightarrow {\rm SU(3)_c \times
SU(2)_L \times U(1)_Y}.
\end{equation} 
In this case the Higgs fields which get VEV at the end inflation
breaking $G_{LR}$ must be in unsafe representations of $G_{GUT}$ in
order to break the $Z_2$ symmetry subgroup of gauge $B-L$ which plays
the role of R-parity \cite{martin}. This happens when the component of
$\Phi$ and $\bar{\Phi}$ transform as an $SU(2)_R$ doublet so that
$\mcN_l =2$ in Eq.~(\ref{loop}).  This includes all the symmetry
breaking pattern of the form
\begin{eqnarray}
{\rm G_{GUT}} &\rightarrow ...\stackrel{\bf Monopoles}\rightarrow & {\rm
SU(3)_c \times SU(2)_L \times SU(2)_R \times U(1)_{B-L}} \nonumber \\
&\stackrel{\bf Inflation +
Embedded \, Strings}\longrightarrow& {\rm SU(3)_c \times SU(2)_L \times
U(1)_Y} \nonumber\\
&\longrightarrow & {\rm SU(3)_c \times U(1)_Q}
\end{eqnarray}
where $G_{GUT} \supset U(1)_{B-L}$. Such models lead to fast proton
decay. This problem can be cured by imposing by hand a discrete
symmetry which plays the role of matter parity. The $B-L$ embedded
strings are unstable \cite{BL}. They can still lead to leptogenesis at
the end of inflation \cite{prl,lepto2,BL}.

\subsection{Non-topologically stable strings}

Topological strings form when $H \rightarrow K$ at the end of
inflation with $\pi_1(H/K) \neq I$. If there is a subsequent phase
transition associate with the breaking $K \rightarrow J$ such that
$\pi_1(G/J) = I$, the strings will rapidly decay after this transition
has taken place. So topological strings can be topologically unstable.
This generically happens when cosmic strings form because K contains a
discrete symmetry which is subsequently broken. In that case the
strings become boundaries of domain walls and the system rapidly
decay. For example, if the breaking at the end of inflation is of the
form
\begin{equation}
H \stackrel{\Phi, \, \bar{\Phi}}{\rightarrow} 
K \times Z_N \rightarrow K,
\end{equation}
then $Z_N$ strings form at the end of inflation and the symmetry
breaking scale is constrained by CMB anisotropies. At the following
phase transition when $Z_N$ breaks down to the identity, the strings
become connected by domain walls and the system rapidly decays
\cite{ShelVil}.

As an example, let's consider the interesting case of unified models
which contain gauged $U(1)_{B-L}$. There is a discrete $Z_2$ symmetry
subgroup of $U(1)_{B-L}$ which can be left unbroken if ``safe''
representations are used to implement the SSB \cite{martin}. If gauged
B-L is broken at the end of inflation with Higgs in safe
representations, the strings which form at the end of inflation are
always topological strings since $\pi_1(H / K \times Z_2)\neq I$. In
this case the component of $\Phi$ and $\bar{\Phi}$ which a acquire a
VEV at the end of inflation transform as an $SU(2)_R$ triplet and
$\mcN_l =3$ in Eq.(\ref{loop}). If unsafe representations are used at
a following intermediate breaking step, the discrete $Z_2$ symmetry will
break and the $Z_2$ strings will rapidly decay. For example,
\begin{eqnarray}
{\rm G_{GUT}} &\rightarrow ... \stackrel{\bf
monopoles}\rightarrow& 
{\rm SU(3)_c \times SU(2)_L \times SU(2)_R \times U(1)_{B-L}} 
\nonumber \\
& \stackrel{\bf inflation + strings}\longrightarrow& 
{\rm SU(3)_c \times SU(2)_L \times U(1)_Y  \times Z_2} \nonumber \\
&\stackrel{\bf strings \, decay}\longrightarrow& {\rm SU(3)_c \times U(1)_Q}.
\end{eqnarray}
Note that if matter parity is broken at low energy, it will have to be
imposed by hand as an additional symmetry.

\subsection{Semilocal strings}

If $H \rightarrow K$ at the end of inflation with a non-minimal set of
Higgs fields, i.e. if there are at least two set of Higgs fields in
the same representation, there is then an extra non Abelian global
symmetry and if $\pi_1 (H/K) \neq I$ semi-local strings form
\cite{semilocal}.

For example, in the case of F-term inflation we can introduce another
pair of Higgs superfields, $\Phi', \bar{\Phi}'$, in the same conjugate
representations as the $\Phi$ and $\bar{\Phi}$ fields in
Eq.~({\ref{W}). The VEV of these fields break $H \rightarrow K$ and we
assume that $\pi_1 (H/K) = Z$. However with two sets of Higgs fields,
the full symmetry is now $H \times SU(2)_{\rm global}$ where $SU(2)$ is a
global symmetry and it is broken down to $K \times U(1)_{\rm global}$. The
vacuum manifold now becomes $S_3$ which does not contain any
non-contractible loops and the strings which form are semilocal
\cite{semilocal}. If the R-charges of these new fields matches those
of the original set of Higgs then
\begin{equation}
W_{\rm infl} = \kappa S (\Phi'\bar{\Phi}' + \Phi\bar{\Phi} - M^2)
\end{equation}
and hence
\begin{equation} V_{\rm infl} = 
\kappa^2 |\bar{\phi} \phi + \bar{\phi}' \phi' - M^2|^2+ \kappa^2
|S|^2(|\bar{\phi}|^2+|\phi|^2 + |\bar{\phi}'|^2+|\phi'|^2 ) + V_D.
\end{equation}
with 
\begin{equation}
V_D = {g^2\over 2}(|\bar{\phi}|^2-|\phi|^2 + |\bar{\phi}'|^2-|\phi'|^2)^2
\end{equation}
The potential has an SU(2) global symmetry and hence the strings
which form at the end of inflation are semilocal \cite{ew}.  

The stability of semilocal strings depends upon the ratio between the
mass of the Higgs field forming the strings and the mass of the string
gauge field $\beta = m_\phi^2/m_A^2$. For $\beta \geq 1$ the strings
are unstable, whereas for $\beta < 1$ the strings are stable
\cite{ShelVil}. Numerical simulations have further shown that when
$\beta \geq 1$, the semilocal strings may not even form
\cite{borrill}. Recent simulations show that deep in the stability
regime, the string contribution to CMB anisotropies is that of global
textures rather than that of topological strings \cite{petja}. From
Ref.\cite{bevis}, we find that $\delta T/T \sim 4 \pi G M^2$ for
global defects, which correspond to $y=2$.  So for the same scale $M$,
strings give a contribution almost $5$ times as small.

D-term strings forming at the end of D-term inflation with to sets of
charged Higgs superfields are BPS states, i.e. $\beta =1$ and they are
unstable \cite{ana}. On the contrary, GUT F-term strings forming at
the end of standard hybrid inflation have $\beta < 1$ \cite{cmb} and
hence promoting them to semilocal strings does not affect their
stability in any essential way.  In that case, the string contribution
to CMB anisotropies is that of global textures rather than that of
topological strings \cite{petja}. P-term F-term strings which form at
the end of F-term P-term inflation correspond to $\kappa = g/\sqrt{2}$
\cite{Kallosh,Pterm}. These strings are BPS states \cite{ana3}, $\beta
=1$ and hence no strings are expected to form. This is particularly
interesting since the parameter space for F-term P-term inflation is
much smaller that GUT F-term inflation \cite{cmb}.

\subsection{Nucleating monopole-antimonopole pairs}

GUT cosmic strings can nucleate monopole-antimonopole pairs and hence
they are not strictly stable. However this is a tunneling process and
therefore the nucleating rate is exponentially suppressed. The
probability for a string to break (per unit length per unit time) is
\cite{Vilenkin82}
\begin{equation}
P \propto \exp(-\pi m^2 / \mu)
\end{equation}
where $m$ is the monopole mass $m \sim 4 \pi M_M/g$ where $M_M$ is the
scale at which the monopoles form; we have $M_{GUT} \gtrsim M_M >
M$. Using Eq.(\ref{mu}) we get
\begin{equation}
P \propto \exp[-N \Big({M_M \over M}\Big)^2].
\end{equation} 
with $N \in [150-150000]$ for $\theta \in [1-10^{-2}]$. Therefore the
strings which form at the end of hybrid inflation are essentially stable.

\section{No strings at all}

If there is no phase transition associated with spontaneous symmetry
breaking at the end or after inflation, apart from the electroweak
phase transition, then there will be no topological defect left in our
visible universe today. This is possible for non minimal models of GUT
hybrid inflation, such as shifted inflation \cite{shifted} or smooth
inflation \cite{smooth}.

\subsection{Shifted inflation}

The superpotential for standard hybrid inflation Eq.~(\ref{W})
contains in principle a infinite serie of non-normalizable terms which
are consistent with the R-symmetry defined below Eq.~(\ref{W}). The
effect of these terms is in the most general case unimportant for the
inflationary dynamics and are usually ignored. However, it was pointed
out in Ref.\cite{shifted}, that if the cut-off scale of the low energy
theory lies between $10^{17}$ GeV and $10^{19}$ GeV \cite{senoguz},
and if the Yukawa coupling of the higher order term is negative and of
order unity, the effect of this latter term on the scalar potential is
crucial. Indeed, for specific values of the parameters, the
inflationary valley of GUT hybrid inflation is then {\em shifted} from
a valley where the VEV of the GUT Higgs fields vanishes to a valley
where it is non-zero and there is spontaneously symmetry
breaking. This is of particular importance for GUT model building,
because one can then break $G_{GUT}$ directly down the standard model,
and keep the unification of the gauge coupling constants' prediction
of the MSSM, and solve the GUT monopole problem at the same time.
Shifted inflation can also occur in models with non minimal content of
GUT Higgs superfields \cite{newshift}.

The superpotential for shifted hybrid inflation is given by 
\begin{equation}
W = \kappa S (\bar{\Phi} \Phi - M^2) - \beta S{(\bar{\Phi} \Phi)^2
\over M_S^2} \label{eq:shifted}
\end{equation}
where S is a gauge singlet and $\bar{\Phi}$ and $\Phi$ are Higgs
superfields in complex conjugate representations of $G_{GUT}$. The
superpotential given in Eq.~(\ref{eq:shifted}) is consistent with an
R-symmetry under which the fields transform as follows: $S
\rightarrow e^{i\theta} S$, $\Phi \rightarrow e^{i\theta} \Phi$,
$\bar{\Phi} \rightarrow e^{-i\theta} \bar{\Phi}$ and $W \rightarrow
e^{i\theta} W$. The SUSY scalar potential which can be calculated from
Eq.~(\ref{eq:shifted}) is
\begin{equation}
V = \left \vert \kappa (\bar{\Phi} \Phi - M^2) -\beta
      \frac{(\bar{\Phi} \Phi)^2}{M_S^2} \right \vert^2 
+ \kappa^2
    \vert S \vert^2 \left(|\Phi|^2 +|\bar{\Phi}|^2\right) \left \vert
      1 - \frac{2 \beta}{\kappa M_S^2} \bar{\Phi} \Phi \right \vert^2
    + {\rm V_D}.
\end{equation}
This scalar potential has a shifted inflationary valley for $1/4 > \xi
> 1/6$, where $\xi = {\beta M^2 \over \kappa M_s^2}$, along which
$S\neq 0$ and $\vert \langle \Phi \rangle\vert = \vert\langle
\bar{\Phi} \rangle\vert = v = ({\kappa M_S^2\over
2\beta})^{1\over2}$\cite{shifted}. We now set $\beta =1$. The common
VEV of the $\Phi$ and $\bar{\Phi}$ fields break the GUT gauge symmetry
during inflation. The scalar potential along the inflationary valley
is given by
\begin{equation}
 V_{\rm infl} = \kappa^2 m^4 + V_{\rm CW} + V_{\rm SUGRA}
\end{equation}
where $m^2 = M_s^2 ({1\over 4 \xi} -1)$. The one-loop corrections
now not only depend on the value of $S$ but also on the common VEV of
the the $\Phi$ and $\bar{\Phi}$ fields, $v$, which is non zero during
inflation \cite{shifted} :
\begin{equation}
V_{\rm CW}= {\kappa^2 m^4 \over 16 \pi^2} \left( 2 \ln
(\frac{\kappa^2 |S|^2}{\Lambda^2} + \ln ((z+1)^2 \ln(1+z^{-1}) +
(z-1)^2 \ln (1-z^{-1}) \right)
\end{equation}
where $z=2 |S|^2/M^2$. Note that $V_{\rm CW}$ does not depend upon the
dimension of the representation of the Higgs fields as opposed to the
standard hybrid case. Incorporating possbile hidden sector VEVs, we
obtain the following SUGRA corrections along the inflationary valley :
\begin{equation}
V_{\rm SUGRA} = \kappa^2 m^4 \left({(|a|^2 + \kappa Ms^2/m_p^2) 
{|S|^2\over m_p^2}} 
+ {|S|^4\over 2 m_p^4} \right) +  \kappa  A m_{3/2} m^2  |S| + \cdots .
\end{equation}
As in the standard hybrid case, shifted inflation is driven by the
loop corrections in most of the parameter space. Inflation ends in a
phase transition when $\eta \simeq 1$ which is very close to the
critical point $S = \sqrt{2} m$. This phase transition is not
associated with spontaneous symmetry breaking. The global SUSY minimum
is at $S=0$ and $v = {1\over 2 \xi} (1 - \sqrt{1 -4\xi})$ and is
constraint by CMB data to be very close to the GUT scale
\cite{senoguz,shifted}. In the case $\beta=1$, when $M_s \sim \mpl$,
shifted inflation can happen for $\kappa \gtrsim 10^{-5}$ whereas if
$M_s \sim 10^{17}$ GeV gives $\kappa \gtrsim 10^{-2}$
\cite{senoguz,shifted}. The constraints on the scale $M$ and on the
spectral index $n_s$ are very similar for shifted inflation and
standard hybrid inflation.  In the case of
shifted inflation, larger values of $\kappa$ are allowed compared with
the standard hybrid case.

\subsection{Smooth inflation}
\label{s:smooth}

Another version of SUSY hybrid inflation which does not lead to the
formation of topological defects at the end of inflation is smooth
inflation \cite{smooth}. The main differences with shifted inflation
is firstly the inflationary valley, where spontaneous symmetry
breaking takes place, which is not flat at tree level. Secondly the
inflationary valley also contains the global SUSY minimum, and hence
there is no ``waterfall'' regime, the inflaton fields smoothly move to
the global SUSY minimum \cite{smooth}. 

The superpotential for smooth inflation is the one used for shifted
inflation without the trilinear term, which is forbidden by imposing
a discrete $Z_2$ symmetry \cite{smooth} :
\begin{equation}
W = \kappa S \( - M^2 + { (\bar{\Phi} \Phi)^2  
\over M_S^2} \). \label{eq:smooth}
\end{equation}
The scalar potential is minimized for $\langle \bar{\Phi}^* \rangle =
\pm \langle \Phi \rangle$. Denoting the common VEV of the real
component of the Higgs field by $\phi$, we have
\begin{equation}
V = (M^4 - {\phi^4\over M^2})^2 + {8 \phi^6 S^2 \over M^4}.
\end{equation}
The parameter space which is compatible with CMB data is very similar
to that of standard hybrid inflation \cite{senoguz,smooth}.

\subsection{Models}

We now discuss various spontaneous symmetry breaking patterns which
are compatible with smooth or shifted inflation as solutions to the
GUT monopole problem. We point out that even though no defect form at
the end of inflation in these cases, for rank greater than 6 gauge
groups, cosmic strings may well form at a subsequent phase transition
and we give further details.

For rank five GUT groups such as ${\rm SU(4)_c \times SU(2)_R \times
SU(2)_L}$ or SO(10) the SSB pattern is as follows
\begin{equation}
{\rm {G}_{GUT}}  
\mathop{\longrightarrow}^{{\rm Shifted \, Inflation}}_{\langle\Phi\rangle, \langle\bar{\Phi}\rangle}{{3_c \,  2_L \, 1_Y}}
\rightarrow{\rm {3_c \, 1_Q}} \label{SSB1}
\end{equation}
and no topological defect form at the end or after inflation. The GUT
monopoles are formed before inflation (alternatively, they may never
form if the GUT symmetry is never restored). 

For GUTs based on a GUT gauge group with rank $>5$ still SSB patterns
of the type Eq.~(\ref{SSB1}) can happen. However, since the rank of the
group is higher, cosmic strings may also form at a following phase
transition. For example,
\begin{equation} 
{\rm G_{GUT}} 
\mathop{\longrightarrow}^{{\rm Shifted \, Inflation}}_{\langle\Phi\rangle, \langle\bar{\Phi}\rangle}\rightarrow {{3_c \, 2_L \, 1 \, 1}} \stackrel{\rm Cosmic \, Strings}\rightarrow{{\large 3_c \, 2_L \, 1_Y}}\rightarrow{{3_c \, 1_Q}}.
\end{equation}
For GUT gauge groups with rank$>6$ there might even be more that on
 type of strings forming after inflation.  For example,
\begin{equation} 
{\rm G_{GUT}} \mathop{\longrightarrow}^{{\rm Shifted \, Inflation}}_{\langle\Phi\rangle, \langle\bar{\Phi}\rangle}\rightarrow {3_c \, 2_L \, 1 \, 1 \, 1} \stackrel{\rm Cosmic \, Strings}\rightarrow {{3_c \, 2_L \, 1 \, 1}} \stackrel{\rm Cosmic \, Strings}\rightarrow
{\rm 3_c \, 2_L \, 1_Y}\rightarrow{\rm 3_c \, 1_Q}.
\end{equation}
This is very generic situation. Note that during inflation the energy
in any hidden sector should be $V_{\rm hid} \lesssim (10^{11}\GeV)^4$
so that the gravitino mass $m_{3/2} \sim V^{1/2}/\mpl \lesssim \TeV$
is sufficiently small.  Larger gravitino masses are incompattible with
the CMB data. This means that any phase transition happining after
inflation should be at sufficiently low scale $\lesssim 10^{11} \GeV$.

\section{Conclusions}

Hybrid inflation is perhaps the best particle physics motivated model
of inflation. It arises naturally in SUSY GUTs \cite{Dvasha,prd} and
it can be viewed as an effective brane inflation action
\cite{Kallosh,Pterm}. It was thus important to take a closer look at the model
and at the parameter space allowed by CMB data. There are two
potential problems. One is that cosmic strings always form at the end
of inflation \cite{prd,new} and their contribution to CMB anisotropies
can be above the 10\% allowed by the data \cite{Pogosian,Fraisse} for
large coupling constants in the F-term case and for gauge coupling $g$
close to the unified coupling in the D-term case \cite{prd,cmb}. The
other is that the scale of inflation may be too high. 

We have shown that there are basically two ways to enlarge the
parameter space for hybrid inflation. The first strategy to enlarge
the parameter space is by looking at and changing the details of the
inflationary scenario itself (not the strings). We found that in
realistic models of GUT F-term inflation, the loop corrections do not
depend on the dimension of the representation of the Higgs fields
which acquire VEV at the end of inflation, but on a restricted number
of component $\mcN_l$ which remain massless during inflation. $\mcN_l$
is found to lie between $1$ and $3$ for phenomenologically interesting
models, and in particular for GUT models which predict massive
neutrinos via see-saw mechanism. 

In the case of D-term inflation, we showed that most of the parameter
space consistent with the data is difficult to reconcile with motivated
values of $g$ when the strings are stable. Furthermore D-term
inflation can occur for values of the inflaton field larger than the
Planck scale; in this case one cannot use minimal K\"ahler potential
to analyse the model. 

Further we showed for both F- and D-term inflation, that the parameter
space could be enlarged by considering decay of the inflaton during
inflation. This requires a large coupling of the Higgs with some other
field.  However, these so called warm inflationary models give a blue
tilted spectral index, at odds with the latest WMAP data.  The bounds
on the inflationary parameters can also be weakend employing the
curvaton and inhomogeneous reheating scenario, or entertaining the
possibility of varying constants.

The second strategy to enlarge the parameter space for hybrid
inflation is to get rid of the strings which form at the end of
inflation or to have no string forming at all. In the latter case, we
need to go beyond standard hybrid inflation, to shifted or smooth
inflation for example \cite{shifted,smooth}. We showed that the
strings may not contribute to CMB anisotropies for a variety of
reasons.  They can be non-topological and unstable such as embedded
strings, they can behave more like global defects and consequently
contribute less to the CMB if they are semilocal, or they can be
topological but not topologically stable.  We explained this in some
detail and gave examples of realistic symmetric breaking patterns,
with particular emphasis on GUTs which predict massive neutrinos via
the see-saw mechanism.

As a final remark, we would like to discuss how hybrid inflationary
confronting WMAP-3 data \cite{WMAP}. WMAP-3 excludes a
Harrison-Zeldovich spectrum at more than three sigma. This excludes
the region where the singlet-Higgs coupling $\kappa$ is smaller than
$10^{-3}-10^{-4}$, see Figs.~\ref{F:F} and \ref{F:D}, in the standard
hybrid model, and $\kappa$ greater than $10^{-6}$ in the case of
varying constants, see Fig.~\ref{F:curv}. If the strings are stable
and contribute to CMB, their contribution may well be too low to be
seen in the CMB temperature anisotropies \cite{cmb}. However, they
could still be detected in upcoming experiments via B-type
polarization of the CMB \cite{Seljak}, since tensor and vector modes
of F-term and D-term strings dominates over the tensor perturbations
of hybrid inflation which are negligible. An appreciable tensor
contribution also shifts the WMAP-3 central value for the spectral
indexto $n_s = 0.98$ \cite{WMAP}. We also note that it is possible to
adjust the spectral index of hybrid inflation by employing non-minimal
K\"ahler potentials, although at the cost of tuning
\cite{Shafilast,McDonald}.  Lastly, WMAP-3 has detected no significant
deviations from a gaussian spectrum.  This may be at odds with hybrid
inflation in the large coupling regime.  As was shown in
Ref.~\cite{Cline} it is possible to generate large second-order
perturbations in CMB due to the instability of the tachyonic field
during preheating, resulting in possibly large non-Gaussianities.

\section*{Acknowledgements}
                                                                               
RJ would like to thank the Netherlands Organisation for Scientific Research
[NWO] for financial support.

\newpage
\appendix
\section{The Barr-Raby model}

In realistic GUT models there will be other GUT Higgs superfields than
the $\Phi_\pm$ fields that couple to the inflaton.  These may be
needed for a variety of reasons: break G down to the intermediate
group, remove pseudo-goldstone bosons from the theory, implement
double-triplet splitting, give masses to the (SM) fermions.  Cross
couplings between the $\Phi,\bar{\Phi}$ and other Higgs fields change
the mass spectrum during inflation, in particular such couplings give
a mass to some components of $\Phi,\bar{\Phi}$.  In this appendix we
give a short overview of the Barr-Raby model, which is a simple and
realistic model to break $SO(10)$ down to the standard model, and
which incorporates hybrid inflation. During inflation 14 out of the 16
component Higgs fields $\Phi$ get massive, and $\mcN \to \mcN_{\rm
light} =2$ in the CMB bounds.

The Barr-Raby model uses an adjoint 45 Higgs $A$, and the complex
conjugate Higgs fields $\Phi$ and $\bar{\Phi}$ in the $16$ and
$\overline{15}$ spinor representation to break SO(10) down to the SM.
In addition there is a pair of spinor $\Phi'$, $\bar{\Phi}'$ Higgses,
2 10-plets $T_1$ and $T_2$, and 4 singlets $S,P,Z_1$ and $Z_2$ whose
roll will become clear below. The superpotential is of the form
\bea
W &\supset& \kappa S(\Phi \bar{\Phi} - \mu^2)+ \frac{\alpha}{4M_s} {\rm tr} A^4
+ \frac12 M_A {\rm tr} A^2 + T_1 A T_2 + M_T T_2^2
\nonumber \\
&+& \bar{\Phi}' \[ \zeta \frac{PA}{M_S} +\zeta_Z Z_1 \] \Phi
+ \bar{\Phi} \[ \xi \frac{PA}{M_S} +\xi_Z Z_2 \] \Phi'
+M_{\Phi'} \bar{\Phi}' \Phi'
\label{WBR}
\eea
The first term in $W$ drives inflation. $S$ is the inflaton which
initially has a large VEV, and inflation is ended by a phase
transition during which the Higgses $\Phi,\bar{\Phi}$ get a VEV in the
$SU(5)$ singlet direction, thereby breaking $SO(10) \to SU(5)$.  The
F-term $F_A$ is minimized by choosing $A$ of the Dimoupoulos-Wilzcek
form: $\langle A \rangle = {\rm diag}(a,a,a,0,0) \otimes i \sigma^2$
with $a = \pm \sqrt{M_A M_S/\alpha}$~\footnote{$F^*_A = $ gives
$\langle A \rangle = {\rm diag}(a_1,a_2,a_3,a_4,a_5) \otimes i
\sigma^2$ with $a_i = a$ or $a_i = 0$. The DW-form is just one of the
many degenerate minima.}.  $\langle A \rangle$ is along the $B-L$
direction; it breaks $SO(10) \to SU(3)_C \times SU(2)_L \times SU(2)_R
\times U(1)_{B-L}$.  The VEV of $\Phi,\bar{\Phi}$ and $A$ together
break $SO(10)$ down to the standard model.

The two SM Higgs doublets correspond to the two doublets in $T_1$ that
remain light after $A$ gets a VEV.  The remaining two triplets in
$T_1$ get superheavy via its coupling to $T_2$~\footnote{$T_2$ is
necessary, since a direct mass term for the triplet components of
$T_1$ would lead to disastrous rapid Higgsino-mediated proton decay}.
This implements the doublet triplet splitting.

The adjoint and spinor Higgs sector need to be coupled to remove
unwanted pseudo goldstone bosons. A direct coupling of the form $W
\supset \bar{\Phi} A \Phi$ would destabilize the doublet-triplet
splitting. Instead the terms in the second line of Eq.~(\ref{WBR}) are
added. Upon minimization of the scalar potential, assuming $\langle P
\rangle$ fixed, the fields $Z_i$ acquire a VEV:
\be
Z_1 = - \frac32 \frac{\zeta}{\zeta_Z} \frac{\langle P \rangle a}{M_s},
\qquad
Z_2 = - \frac32 \frac{\xi}{\xi_Z} \frac{\langle P \rangle a}{M_s}
\ee
Plugging back the VEV into the superpotential the mass terms in the
spinor sector are of the form
\be
W \supset
\kappa S ( \bar{\Phi} \Phi - M^2)+
\bar{\Phi}' \[ \frac{\zeta P a}{M_s} \frac32(B-L-1) \] \Phi
+ \bar{\Phi} \[ \frac{\zeta P a}{M_s} \frac32(B-L-1) \] \Phi'
+M_{\Phi'} \bar{\Phi}' \Phi'
\ee
We label the Higgses in analogy with the fermion masses, by their SM
quantum numbers:
\bea
16 &=& (1,1,0) + (\bar{3},1,1/3) + (1,2,-1/2) + (\bar{3},1,-2/3) + (3,2,1/6) 
+(1,1,1)
\nonumber\\
&=& \phi_{\bar{\nu}} + \phi_{\bar{d}} + \phi_L + \phi_{\bar{u}} + \phi_Q 
+ \phi_{\bar{e}}
\eea
and similarly for the barred and primed fields.  The mass term is
diagonal in that the it only couples SM plets of the same or conjugate
type.

Since $\frac32(B-L -1)=0$ for both the ``RH neutrino'' and
``positron'' components the superpotential in these sectors simply
read
\be
W \supset \kappa S  \bar{\phi}_i \phi_i
\hspace{2cm}{\rm for} \; i = \nu^c, e^c
\ee
Hence the ``RH neutrino'' and ``positron'' of $\bar{\Phi},\Phi$ remain
light during inflation; they do not get an extra contribution to their
mass due to the coupling with other Higgses.  All other components
have $\frac32(B-L -1)\neq0$, and become heavy during inflation.  For
example, the superpotential in the $Q$ sector reads
\be
W = \kappa S \bar{\phi}_Q \phi_Q
- M1 \bar{\phi}'_Q \phi_Q
- M2 \bar{\phi}_Q \phi'_Q
+M_{\Phi'} \bar{\phi}'_Q \phi'_Q
\ee
with $M_1 = \frac{\zeta P a}{M_s}$ and $M_2 = \frac{\zeta P a}{M_s}$,
which gives a heavy mass to the fields in this sector (assuming $M_i
\sim M_{\rm GUT} > \kappa M$ with $\kappa M$ setting the scale of
inflation).

The important point for hybrid inflation is that there are only two
light Higgs components during inflation (the RH neutrino and positron
component).  Only these two fields contribute to the Coleman-Weinberg
potential and thus the effective dimensionality, which equals the
number of light fields, is $\mcN_l = 2$ instead of $\mcN = 16$.  The
story above goes also throught if we introduce Higgs fields
$\Phi,\bar{\Phi}$ and $\Phi',\bar{\Phi}'$ in the $126$ and
$\overline{126}$ representation.  After the coupling with the adjoint
and singlet fields only an $SU(2)_R$ triplet remains massless, and
thus in this case $\mcN_l =3$.

\end{document}